\documentclass[amsfonts, amssymb, amsmath, twocolumn, showkeys, nofootinbib,superscriptaddress, twoside, aps, prx]{revtex4-2}
\usepackage[english]{babel}
\usepackage[utf8]{inputenc}
\usepackage[colorinlistoftodos, color=green!40, prependcaption]{todonotes}
\usepackage{tabularx}
\usepackage{amsthm}
\usepackage{mathtools}
\usepackage{physics}
\usepackage{xcolor, color}
\usepackage{graphicx}
\usepackage{adjustbox}
\usepackage{placeins}
\usepackage[T1]{fontenc}
\usepackage{lipsum}
\usepackage{csquotes}
\usepackage{svg}
\usepackage[pdftex, pdftitle={Article}, pdfauthor={Author}]{hyperref} 
\usepackage{soul}
\usepackage{multirow}

\begin{document}

\title{State-Insensitive Trapping of Alkaline-Earth Atoms in a \protect\\Nanofiber-Based Optical Dipole Trap}
    
\author{G. Kestler}
    \affiliation{Department of Physics, University of California San Diego, California 92093, USA}

\author{K. Ton}
    \affiliation{Department of Physics, University of California San Diego, California 92093, USA}

\author{D. Filin}
    \affiliation{Department of Physics and Astronomy, University of Delaware, Newark, Delaware 19716, USA}

\author{C. Cheung}
    \affiliation{Department of Physics and Astronomy, University of Delaware, Newark, Delaware 19716, USA}

\author{P. Schneeweiss}
    \affiliation{Department of Physics, Humboldt Universit\"at zu Berlin, 12489 Berlin, Germany}
    
\author{T. Hoinkes}
    \affiliation{Department of Physics, Humboldt Universit\"at zu Berlin, 12489 Berlin, Germany}
    
\author{J. Volz}
    \affiliation{Department of Physics, Humboldt Universit\"at zu Berlin, 12489 Berlin, Germany}
    
\author{M. S. Safronova}
    \affiliation{Department of Physics and Astronomy, University of Delaware, Newark, Delaware 19716, USA}
       
\author{A. Rauschenbeutel}
    \affiliation{Department of Physics, Humboldt Universit\"at zu Berlin, 12489 Berlin, Germany}
    
\author{J. T. Barreiro}
    \affiliation{Department of Physics, University of California San Diego, California 92093, USA}
    
\begin{abstract}

Neutral atoms that are optically trapped using the evanescent fields surrounding optical nanofibers are a promising platform for developing quantum technologies and exploring fundamental science, such as quantum networks and many-body physics of interacting photons. Building on the successful advancements with trapped alkali atoms, here we trap strontium-88 atoms, an alkaline-earth element, in a state-insensitive, nanofiber-based optical dipole trap using the evanescent fields of an optical nanofiber. Employing a two-color, double magic-wavelength trapping scheme, we realize state-insensitive trapping of the atoms for the kilohertz-wide $5s^{2}\;^{1}\!S_{0}-5s5p\;^{3}\!P_{1,|m|=1}$ intercombination transition, which we verify by performing high-resolution spectroscopy for an atom-surface distance of only $\approx$300~nm. This allows us to experimentally find and verify the state-insensitivity of the trap nearby a theoretically predicted magic wavelength of 435.827(25)~nm, a necessary step to confirm precision atomic physics calculations. Alkaline-earth atoms also exhibit non-magnetic ground states and ultra-narrow linewidth transitions making them ideal candidates for atomic clocks and precision metrology applications, especially with state insensitive traps. Additionally, given the low collisional scattering length specific to strontium-88, this work also lays the foundation for developing versatile and robust matter-wave atomtronic circuits over nanophotonic waveguides.

\end{abstract}

\maketitle

\section{Introduction}
Hybrid quantum systems with cold atoms coupled to photonic devices are a strong candidate for building {\em atomtronic} circuits; quantum circuits analogous to electronic circuits based on atomic matter waves instead of electrons~\cite{Seaman2007, Amico2022}. For example, cold atoms trapped in free-space optical potentials have already realized integrated atomtronic elements~\cite{Stickney2007, Pepino2009, Ryu2015, Wang2015, Caliga2017}, and similarly, ring-shaped Bose-Einstein condensates have been tailored for quantum sensing~\cite{Navez2016, Kumar2016, Bell2016, Aghamalyan2016}. However, photonic integrated electrical and optical circuits that can realize near-surface optical and magnetic trapping potentials offer more compact and diverse geometries for quantum many-body simulations~\cite{Douglas2015, Hung2016}, cavity quantum electrodynamics (QED)~\cite{Yalla2014, Schneeweiss2016, Chang2019, Chang2020}, and to coherently control and manipulate matter waves~\cite{Wang2005, Chang2018, Ovchinnikov2022}.

In this context, tapered optical fibers with a sub-wavelength diameter, ``nanofiber'' waist have proven to be a versatile platform for trapping and optically interfacing laser-cooled atoms using the evanescent field surrounding the nanofiber~\cite{Kien2004, Kien2004a}. In the last decade, such nanofiber-based optical interfaces have been demonstrated with two types of alkali atoms, cesium~\cite{Vetsch2010a, Goban2012a, Beguin2014, Kato2015} and rubidium~\cite{Lee2015, Gokhroo2022}, and have contributed to advancing quantum science and technology~\cite{Corzo2019}, such as chiral quantum optics~\cite{Petersen2014, Lodahl2017} and quantum memories of light~\cite{Patnaik2002, Gouraud2015, Sayrin2015}.

Notwithstanding these achievements, alkaline-earth atoms, such as strontium, would offer several unique advantages over alkalies when interfaced with nanofibers. Using the $5s^{2}\;^{1}\!S_{0}-5s5p\;^{3}\!P_{1}$ (henceforth ${}^{1}\!S_{0}-{}^{3}\!P_{1}$) intercombination transition for implementing a magneto-optical trap (MOT) with $^{88}$Sr, atomic clouds can reach $\approx$1~$\mu$K temperatures without the need for additional laser cooling, thus reducing the potential depths necessary to load atoms into evanescent field traps from a MOT. This narrow 7.6-kHz transition also enables high-resolution spectroscopy for investigating atom-surface interactions~\cite{Martin2017, Cook2017}. In addition, the spherically symmetric, non-magnetic ground state of ${}^{88}$Sr reduces sensitivity to magnetic field noise, and the small collisional scattering length minimizes decoherence in matter-wave applications~\cite{Escobar2008, Aguila2018}. Experiments in chiral quantum optics will also benefit from the polarization-maintaining nature of strontium atoms as scatterers, allowing one to tune the directionality of scattering into the guided mode of the waveguide via polarization of the excitation light field~\cite{Petersen2014, Mitsch2014, Guimond2016, Lodahl2017, Pucher2022}. In conjunction with the propagation direction-dependent polarization of the nanofiber-guided light, one can, for example, study the dependence of collective radiative phenomena on the directionality of the emission.

Laser-cooling and precision-spectroscopy scenarios involving narrow-line atomic transitions demand minimal differential energy AC-Stark shifts. This is particularly relevant for two-color trapping schemes, which rely on opposing optical forces, and thus involve higher optical intensities than free-space optical dipole traps to reach the same trap depths. Furthermore, the trapping fields are highly non-uniform near the surface of nanophonic devices, thus inducing strong spatially dependent light shifts of the atomic transition across the trapping potential~\cite{McKeever2003, Kien2004, Chang2019}. These issues can be resolved when using {\em magic} wavelengths for which the atomic polarizabilities are identical for the upper and lower level of the corresponding transition, thus providing a state-insensitive trap at all locations. In free-space traps, the use of magic wavelengths is an effective tool for atomic time keeping~\cite{Ludlow2006, Boyd2007, Bloom2014}, precision spectroscopy~\cite{Ye2008, Okaba2014}, quantum computation~\cite{Safronova2003, Saffman2005, Saffman2016, Weiss2017}, and in tightly focused ``optical tweezers'' in particular~\cite{Hutzler2017, Cooper2018, Covey2019}. In evanescent field traps specifically, an absence of magic wavelengths leads to spatially dependent light shifts on the MOT cooling transition, which in turn severely complicate efficient loading of atoms from the MOT into the evanescent fields of photonic integrated circuits, nanofibers and optical cavities~\cite{McKeever2003, Kien2004}.

Here, we report on trapping of strontium atoms in the evanescent fields surrounding an optical nanofiber. The two-color optical dipole trap confines 61(7) atoms for an average lifetime of $\tau=7.4(7)$~ms. Using a measurement based on a theoretical magic wavelength of the ${}^{1}\!S_{0} - {}^{3}\!P_{1}$ transition at 435.827(25)~nm, computed in this work, we also experimentally verify state-insensitive trapping.

This work also provides a much needed benchmark of high-precision theory. Accurate computation of atomic polarizabilities of strontium's excited states, needed for the prediction of magic wavelengths, becomes increasingly difficult for shorter wavelengths. The main contributions to the $^3\!P_1$ polarizability at 435.8~nm come from the transition to the rather high $5s7s$~$^{3}\!S_{1}$ state. There are no precision experimental benchmarks for such transitions. Moreover, the number of contributing resonances rapidly increases as a result of the increased density of states involving higher principal quantum numbers. The experimental agreement with the theoretical prediction of this magic wavelength validates the methodology of computing the polarizabilities of atoms with a few valence electrons, as well as our methods to assess the accuracy of the theory. Reliable computations of polarizabilities at shorter wavelengths are needed for the many cold atom applications mentioned above.

\section{Experimental Setup}
\label{sec:exp_setup}

\begin{figure}[t!]
    \includegraphics{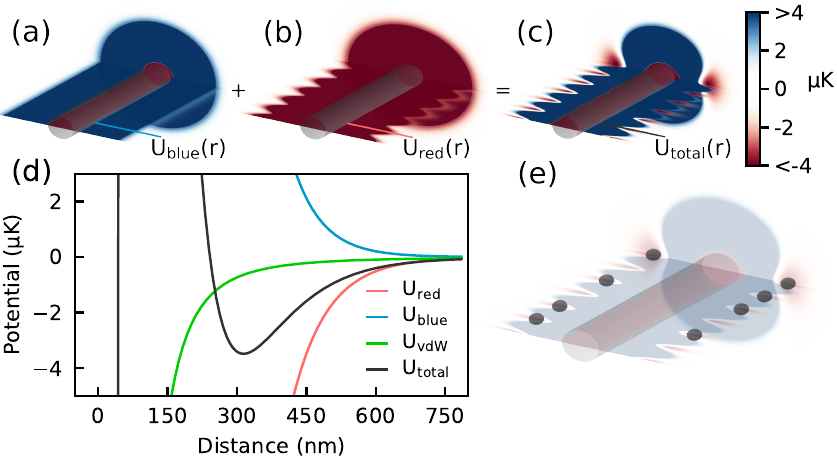}
    \caption{The two-color evanescent field trapping potential. (a) A repulsive blue-detuned potential that is combined with (b) an attractive, red-detuned lattice results in (c) a three dimensional optical dipole trap around the nanofiber. (d) Total potential of the nanofiber-based trap (black curve) along the \emph{radial} direction as a function of distance from the nanofiber's surface and arising from the sum of optical potentials (red and blue) and van der Waals potential of the nanofiber (green). (e) Individually trapped strontium atoms (dark grey spheres) are confined radially, azimuthally, and axially along the nanofiber by the potential illustrated in (c).}
  \label{fig:potentials}
\end{figure}

In this work, strontium atoms are trapped along the length of the nanofiber at a distance of $\approx$300~nm using an optical potential that is generated by the evanescent fields of two nanofiber-guided light fields [see Fig.~\ref{fig:potentials}]. The first trapping field ($\lambda\approx436$~nm), which is blue-detuned from the strong 461-nm dipole transition, produces a second optical potential that repels atoms from the nanofiber surface [see Fig.~\ref{fig:potentials}(a)]. An additional trapping field forms a standing wave along the fiber waist and has a free-space wavelength of $\approx$473~nm, which is red-detuned from the strongest dipole transition, attracting the atoms towards the nanofiber surface [see Fig.~\ref{fig:potentials}(b)]. The depth of the repulsive (attractive) trapping potential is proportional to the light's intensity and the atomic polarizability at the trapping field wavelength of 436~nm (473~nm), the latter being large ($\approx$-1500 (3500) a.u.) because of the proximity to the strong 461-nm dipole transition.

Because the intensity of the blue-detuned evanescent field has a smaller decay length than the red-detuned field, the combination of the corresponding potentials has a local minimum which, for our settings, is at a distance of about 300~nm from the fiber surface providing \emph{radial} confinement for the atoms, even in the presence of the van der Waals and Casimir-Polder surface potentials [see Fig.~\ref{fig:potentials}(c-d)]. For our resonant wavelength of $\lambda=461$~nm, the van der Waals potential dominates at very close distances $\lesssim\lambda/2\pi$ while the Casimir-Polder potential is more accurate for distances $\gtrsim\lambda$~\cite{Casimir_Polder1948, Sukenik1993, Landragin1996, Garcion2021}. Thus, our nanofiber-based trap operates in the cross-over region between these two.  We note that the atom-fiber distance can be tuned by changing the ratio of the optical powers of the trapping fields. \emph{Azimuthal} confinement around the fiber is achieved by choosing orthogonal quasi-linear polarizations for the two trapping fields~\cite{Vetsch2010a} while \emph{axial} confinement along the fiber is realized by launching two counter-propagating red-detuned fields through the nanofiber, thereby forming an optical lattice along the fiber waist [see Fig.~\ref{fig:potentials}(e)] (see Appendix~\ref{app:trap_potentials}).

The required evanescent trapping fields at these wavelengths are made possible by tapering the fiber with a flame pulling technique to a nominal waist diameter of 230~nm~\cite{Stiebeiner2010}. We choose this diameter to optimize the atom-photon coupling at our probing wavelength of 461~nm, which occurs at a waist of $\approx$$\lambda/4$~\cite{Warken2007, Stiebeiner2010}. Furthermore, the fiber only guides the fundamental $\rm{HE}_{11}$ mode for all wavelengths of interest in the {400--500~nm} range. This nanofiber is first loaded into an ultra-high vacuum load-lock chamber before insertion into our main science chamber, which is held at a pressure in the $10^{-11}$~Torr range throughout the experiment~\cite{Kestler2019}.

\begin{figure}[t!]
    \includegraphics[width=\columnwidth]{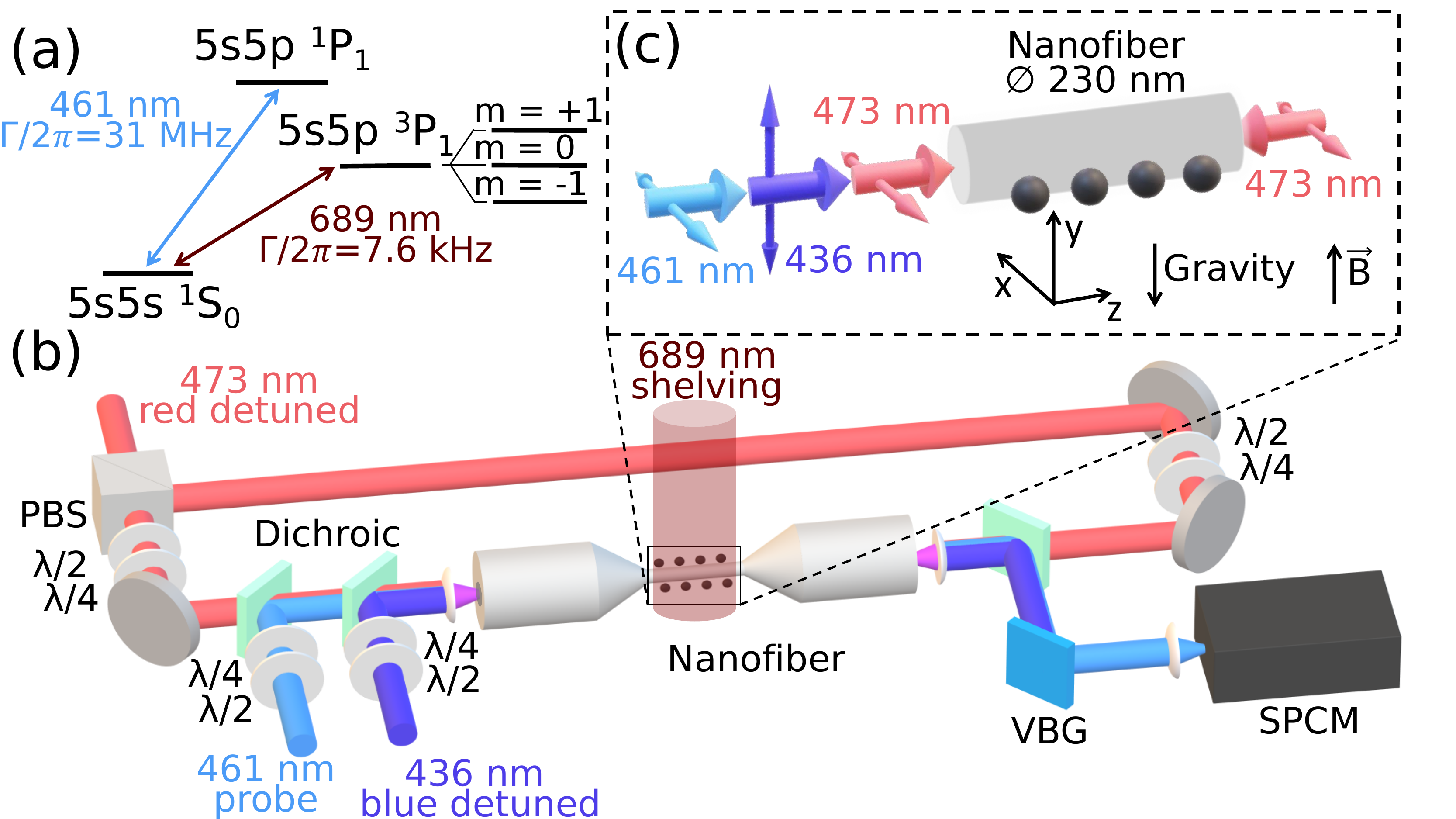}
    \caption{(a) Strontium energy levels relevant to the experiment. (b) Schematic of the experimental setup for the state-insensitive nanofiber-based trap for strontium atoms (dark grey spheres). The polarization of each beam is individually controlled by a set of half-wave ($\lambda/2$) and quarter-wave ($\lambda/4$) plates. PBS: polarizing beam splitter, VBG: volume Bragg grating, SPCM: single-photon counting module. (c) Propagation direction (thick arrows) and polarization (thin arrows) of the different nanofiber-guided fields necessary for trapping potentials in all three dimensions [see Section~\ref{sec:exp_setup}]}.
  \label{fig:setup}
\end{figure}

Our experimental cycle starts by generating an ultracold MOT cloud of ${}^{88}$Sr atoms, which we overlap with the waist region of the fiber, which is 5~mm long. This ultracold atomic cloud has a free-space optical depth (OD) of 3.5 and reaches temperatures of $\approx$1~$\mu$K after MOT-cooling on the 689-nm transition~\cite{Campbell2017, Nicholson2015}. The fiber-guided trapping fields are then switched on with the MOT present for 35~ms to load the trap with cooled atoms before the MOT beams are switched off. The trapped atoms are ultimately probed by measuring absorption on the strong 461-nm dipole transition through the fiber [see Fig.~\ref{fig:setup}(a-b)]. We are unable to probe on the narrow 689-nm transition because the high bending losses in our current setup prevent us from sending the light through the nanofiber. Thus, instead, we employ an external 689-nm shelving beam to characterize and measure the novel magic wavelength [see Fig.~\ref{fig:setup}(b)]. In all experiments, to ensure spin-resolved spectroscopy, the quadrupole B-field is kept on for the duration of the experimental cycle and contributes a $\approx$200~mG magnetic field in the vertical direction at the atoms.

\section{Trapping of Strontium}

From the ultracold atomic cloud, we end up trapping 61(7) atoms in a two-color, nanofiber-based optical dipole trap for an average lifetime of $7.4(7)$~ms. Using our reported magic-wavelength for the red-detuned trapping field at a free-space wavelength of 473.251~nm~\cite{Kestler2022}, and a blue-detuned trapping field at a free-space wavelength of 435.97~nm, we realize the optical potential shown in Fig.~\ref{fig:potentials} with a calculated depth of $3.5(3)$~$\mu$K at 320(10)~nm from the nanofiber surface, with uncertainties stemming from the error of experimentally determining the intensities and polarizations of the nanofiber-guided trapping fields (see Appendix~\ref{app:pol_calibration}). This total potential also includes a contribution from the van der Waals and Casimir-Polder potentials [see Fig.~\ref{fig:potentials}(d)]. For these trap parameters, we use a power of 1.03(1)~mW for the running-wave, blue-detuned trapping field, and $2\times32(2)$~$\mu$W for the standing-wave, red-detuned trapping field (see Appendix~\ref{app:power_calibration}).

\begin{figure}[t!]
    \includegraphics{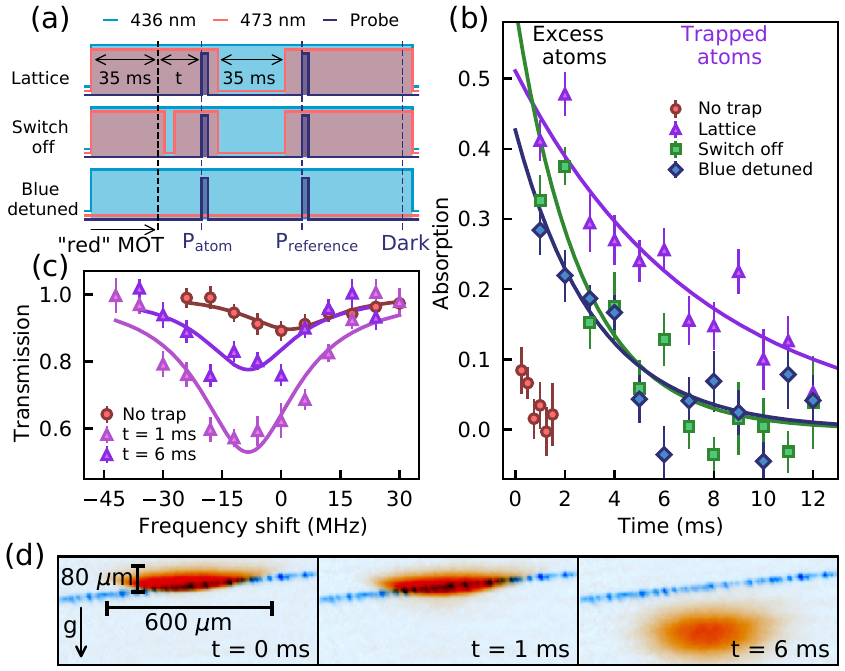}
    \caption{(a) Experimental sequences indicating the on and off times of the red- and blue-detuned trapping fields at $\approx$473~nm and $\approx$436~nm, respectively, as well as the timing of the 461-nm probe pulses. In the following, the three settings (top to bottom) are referred to as ``lattice'', ``switch off'', and ``blue detuned'', respectively. See main text for details.(b) Absorption of the nanofiber-guided, 461-nm probe field as a function of waiting time for four settings: without trapping fields (``no trap'' reference measurement) and for the three settings shown in (a). The waiting time $t$ is measured relative to the turning off of the MOT beams. (c) Transmission spectroscopy on the probing transition for the ``lattice'' trapping setting with ${t = 1}$~ms and ${t = 6}$~ms as well as for the ``no trap'' setting. The spectral shapes are fit to a Lorentzian. (d) Absorption image of the MOT cloud (red) overlaid with the nanofiber's scattering image (blue) at different times after turning off the MOT beams. The atomic cloud expands slowly while falling under gravity. At ${t = 6}$~ms, the atomic cloud no longer overlaps with the nanofiber.}
    \label{fig:state_insensitive_trapping}
\end{figure}

In order to verify the trapping of atoms, we measure the absorption of a nanofiber-guided probe field resonant with the $5s^{2}\;^{1}\!S_{0} - 5s5p\;^{1}\!P_{1}$ dipolar transition with a free-space wavelength of $\approx$461-nm (henceforth $^{1}\!S_{0} - ^{1}\!P_{1}$). If atoms are located in the evanescent field surrounding the nanofiber, a fraction of this probe field will be absorbed. For all absorption measurements in this work, the plane of quasi-linear polarization of the probe field coincides with that of the red-detuned trapping light field [see Fig.~\ref{fig:setup}(c)]. We use a single-photon counting module (SPCM) to record the probe power at the output of the nanofiber [see Fig.~\ref{fig:setup}(b)]. The top panel in Fig.~\ref{fig:state_insensitive_trapping}(a) shows the experimental sequence for measuring the absorption of the probe field by the atoms near the nanofiber. After switching off the MOT beams, we wait for a variable time $t$ and then send a first probe pulse. After transmission through the atoms along the nanofiber, we measure its power with an SPCM. We then switch off the red-detuned trapping fields for 35~ms such that atoms are expelled from the trap. After switching the red-detuned trapping fields on again, we then send a second probe pulse that serves as a reference. The two probe pulses have the same power of 50(3)~pW in the nanofiber and the boxcar-shaped pulses have a duration of 40~$\mu$s.

Figure~\ref{fig:state_insensitive_trapping}(b) shows the measured absorption of the probe field for waiting times ranging from $t=1$~\rm{ms} to 12~ms (purple triangles). Fitting an exponential decay curve to the data yields a $1/e$ decay time of $\tau=7.4(7)$~ms (purple line). This result is short compared to equivalent free-space optical dipole traps, which could be due to the fact that nanofiber-based traps are known to incur additional heating due to phonon-photon coupling between the mechanical structure and trapping fields~\cite{Hummer2019} (see Appendix~\ref{app:heating_rates}). For comparison, we measure the absorption of the probe field when no trapping fields are present in the nanofiber (red open circles). We also measure the absorption of the probe field when only the blue-detuned trapping field is present in the fiber (blue diamonds). In this case, we still observe an absorption signal with a decay constant of $\tau_{\rm{bd}}=3.2(4)$~ms. To verify that the measured absorption signal in the ``lattice'' setting is indeed a result of trapped atoms, we switch off the red-detuned trapping field for 200~$\mu$s at the beginning of the lifetime measurement, $\approx$1~ms after turning off the MOT beams  (green squares). In this ``switch-off'' setting, the absorption signal has a decay constant of $\tau_{\rm{so}} = 2.7(3)$~ms, which is comparable to the decay constant of the absorption signal that we observe in the ``blue-detuned'' setting. This confirms that in the ``lattice'' setting, atoms are indeed trapped in the nanofiber-guided trapping fields. As discussed in detail in Appendix~\ref{app:blue_detuned_density_enhancement}, the presence of a blue-detuned potential enhances the atom density near the fiber. This explains the differences in initial absorption signal between the "blue-detuned"/"switch-off" and the "no trap" scenarios. We expect the decay constant of the latter scenario to be the same as the former two, however, the atom density around the fiber is quickly at the noise level without trapping fields to confirm this experimentally. 

We now estimate the number of atoms loaded into our trap by fitting a Lorentzian model to the absorption profile of the nanofiber-trapped atoms~\cite{Vetsch2010a}. The choice for a Lorentzian fit is because we do not expect significant broadening of the $\approx$30~MHz transition. In order to avoid the effect of untrapped atoms from the falling MOT cloud, we record the absorption profile after holding the atoms in the trap for 6~ms [see purple triangles in Fig.~\ref{fig:state_insensitive_trapping}(c)]. At this moment, the MOT cloud is no longer overlapped with the fiber [see Fig.~\ref{fig:state_insensitive_trapping}(d)] and the maximal OD of the trapped atoms is 0.25(3), which corresponds to 27(3) atoms. Given the exponential lifetime of the trapped atoms, we can infer that 61(7) atoms would have been trapped immediately after turning off the MOT beams (see Appendix~\ref{app:evanescent_fields_number}).

As shown in Fig.~\ref{fig:state_insensitive_trapping}(c), the Lorentzian fit (purple line) to the data taken after holding the atoms for 6~ms yields a shift of the $^{1}\!S_{0} - ^{1}\!P_{1}$ resonance, inferred to be of $\Delta=-8.3(1.7)$~MHz, which we confirm by performing an additional spectroscopy scan after holding the atoms for 1~ms when the absorption signal is stronger (pink triangles and line) (see Appendix~\ref{app:blue_detuned_light_shift}). This shift is presumed to be a result of a large AC-Stark shift of the $^{1}\!P_{1}$ state. The magic wavelength in this paper is with regards to the $^{1}\!S_{0} - ^{3}\!P_{1}$ transition and it is unaffected by the $-8.3(1.7)$~MHz shift. To account for this light shift, we detune our 461-nm probing laser by this $\Delta$ for all absorption measurements. 

\section{Magic Wavelength of Strontium}
\label{sec:magic_theory}

For realizing a state-insensitive nanofiber-based atom-trap, magic wavelengths have to be employed for red- and blue-detuned trapping fields, as has been demonstrated for the case of Cs and its D2-line transition~\cite{Lacroute2012, Goban2012a}. For the $^{1}S_0  - ^{3}P_{1, |m|=1}$ transition in $^{88}$Sr, a red-detuned magic wavelength has previously been determined~\cite{Kestler2022}. However, a blue-detuned magic wavelength still has to be found. This requires accurately calculating the atomic polarizabilities of both the ground and excited states to determine a wavelength below 461~nm where they are equal. The formalism for these calculations was described in Ref.~\cite{Kestler2022}. Therefore, we focus on the specific challenges of computing this magic wavelength and provide further computational details in Appendix~\ref{app:blue_magic_theory}.
 
While the polarizability has the scalar, vector, and tensor components ($\alpha_{0}, \alpha_{1}$, and $\alpha_{2}$ respectively), only the scalar component is non-zero for the spherically symmetric ground state with zero angular momentum. For the $5s5p\;^{3}\!P_{1}$ state, the vector term vanishes for the case of purely linear polarization, which prevails for the red- and blue-detuned trapping field and the position of the potential minimum. Thus, the polarizability of the $5s5p\;^{3}\!P_{1}$ state depends only on the scalar and tensor parts in this case \cite{Zheng2020}:
\begin{equation}
    \label{eq:pol}
    \alpha = \alpha_{0} + \alpha_{2}\bigg(\frac{3\cos^{2}\theta_{p} - 1}{2}\bigg)\frac{3m^{2} - J(J+1)}{J(2J-1)}.
\end{equation}
Here, $\theta_{p}$ denotes the angle between the linear polarization and magnetic quantization axis and we consider the case for $\theta_{p}=0$ and $|m|=1$, so that $\alpha=\alpha_0+\alpha_2$.
The magic wavelength is determined as the one where the $^1\!S_0$ and $^3\!P_{1,|m|=1}$ polarizability curves cross.

\begin{figure}[t!]
    \includegraphics{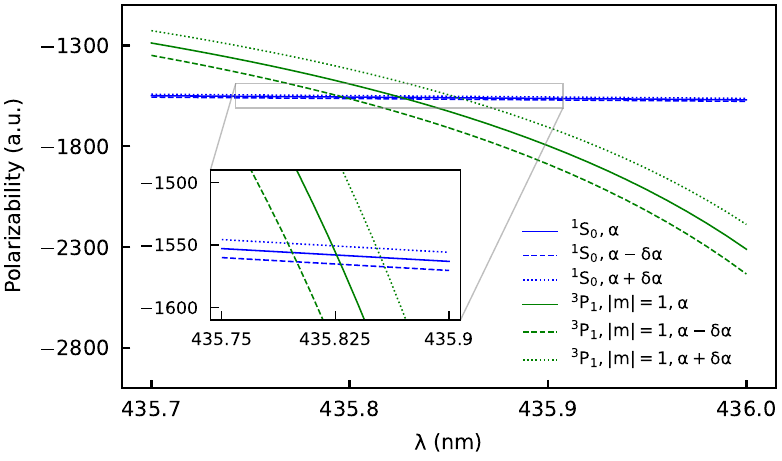}
    \caption{Theoretical calculation of the magic wavelength at 435.827(25)~nm for the $5s^2\,^1\!S_0 - 5s5p\,^3\!P_1$ transition.}
    \label{fig:theory436}
\end{figure}

We use a hybrid method that combines  the configuration interaction (CI) and the linearized coupled-cluster approaches  (refereed to as the CI + all order method ~\cite{2009}) to compute Sr matrix elements and the polarizabilities. 
The results of the calculation of the magic wavelength for the $^1\!S_0 - ^3\!P_1$ transition near 436 nm are illustrated in  Fig.~\ref{fig:theory436}.
The magic wavelength is obtained as the crossing point of the $^1\!S_0$ and the $^3\!P_{1,|m|=1}$ curves. The $^1\!S_0$ polarizability does not have any resonances (i.e., allowed transitions from $^1\!S_0$ to other states) in the vicinity of the magic wavelength and its curve is essentially flat. 

The closest $^3\!P_1$ resonance to the magic wavelength is the 
$5s5p \,^3\!P_1 - 5s7s\, ^3\!S_1$ transition at 436.3~nm. This transition contributes 81\% to the $^3\!P_{1,|m|=1}$ polarizability value. The uncertainty in this matrix element dominates the uncertainty of the polarizability $\delta\alpha$ at the magic wavelength, with all other uncertainty sources being negligible. We plot $\alpha+\delta \alpha$ and $\alpha-\delta \alpha$ curves in Fig.~\ref{fig:theory436}. The uncertainty in the value of the magic wavelength is determined from the crossings of these additional curves. 

The challenge of the present calculation was to establish the numerical stability of the computation of such a highly excited state, beyond the scope of our previous applications of the CI+all-order approach. In our calculations, as it is standard in order to replace the bound state spectrum + continuum by a finite basis set, we create basis set orbitals in a spherical cavity of radius $R$, where the cavity size has to be large enough to accommodate the $7s$ orbital. To ensure accurate description of the $5s7s$ state, we carried out computations of a significantly large number of states and compared the resulting energies with the NIST database \cite{NIST}. We find that the relative accuracy of the energies remains about the same until we reach $5s8d$ and $5s10s$ states, at which point it starts to decrease. The energy of the $5s7s\, ^3\!S_1$
state differs from the experiment reported in the NIST database~\cite{NIST} by 0.46\% and the $5s5p \,^3\!P_1 - 5s7s\, ^3\!S_1$ transition energy by only 0.24\%. We have also performed several calculations with different cavity sizes, from $R=60$~a.u. to $R=80$~a.u. and found less than 0.01\% changes in the $5s7s$ energy. Therefore, we do not expect any additional uncertainty associated with the higher energy of the $5s7s\, ^3\!S_1$ state.

The same methodology can in principle be used to theoretically verify the $\Delta=-8.3(1.7)$~MHz shift of the $^{1}\!S_{0}-{}^{1}\!P_{1}$ transition in Fig.~\ref{fig:state_insensitive_trapping}(c). Unfortunately, the computation of the  polarizabilities of the $^1P_1$ state, that has much higher energy, is far more challenging at short wavelengths, since dominant contributions to its polarizability at 435.8~nm come from even higher states, such as $5s11d$. To properly describe $5s11d$ states, the cavity must be large enough to contain $11d$ orbitals leading to numerical instabilities with cavities sizes over 80~a.u. The exploratory computations carried out in this work identified the need to develop better basis sets and grid distributions that allow one to accurately describe these highly excited states, which would be a subject of future work. This problem appears to be specific to the configuration interaction approach that we use to correlate valence electrons as we were able to successfully compute polarizabilities of the 18s states of the monovalent Rb \cite{18s}.

\section{State-insensitive trapping of strontium}

In order to check that we indeed realized a state-insensitive trap for the ${}^{1}S_{0}-{}^{3}P_{1}$ transition, we perform a shelving-type spectroscopy on the near-surface trapped atoms. In principle, there are 3 contributions that can give rise to a shift of the corresponding atomic resonance: the surface-force induced shifts as well as the shifts from the red- and blue-detuned trapping fields. We eliminate the effect of the attractive red-detuned field by choosing the experimentally determined magic wavelength of the $\,^1\!S_0$ and $\,^3\!P_1$ states as described in Ref.~\cite{Kestler2022}. In our nanofiber trap, the red-detuned trapping field is horizontally polarized and the magnetic quantization axis from the quadrupole B-field is aligned vertically, which yields a magic wavelength of 473.251~nm. Furthermore, at the trap distance $>300$~nm from the fiber surface, the Casimir-Polder potential is much smaller than the width of the resonance, and thus does not induce a detectable shift. The only remaining contribution to energy shifts in our trap configuration then stems from the blue-detuned trapping field.

While we could in principle directly probe the 689-nm atomic transition using fiber-guided light fields, we use an external laser beam as our current setup suffers from high bending losses at 689~nm in the fiber tails on either side of the tapered nanofiber. We therefore rather send a free space (non-guided) beam onto the fiber tapered region tuned to the ${}^{1}S_{0} - {}^{3}P_{1,m=-1}$ resonance~[see Fig.~\ref{fig:setup}(b)].

\begin{figure}[t!]
    \includegraphics{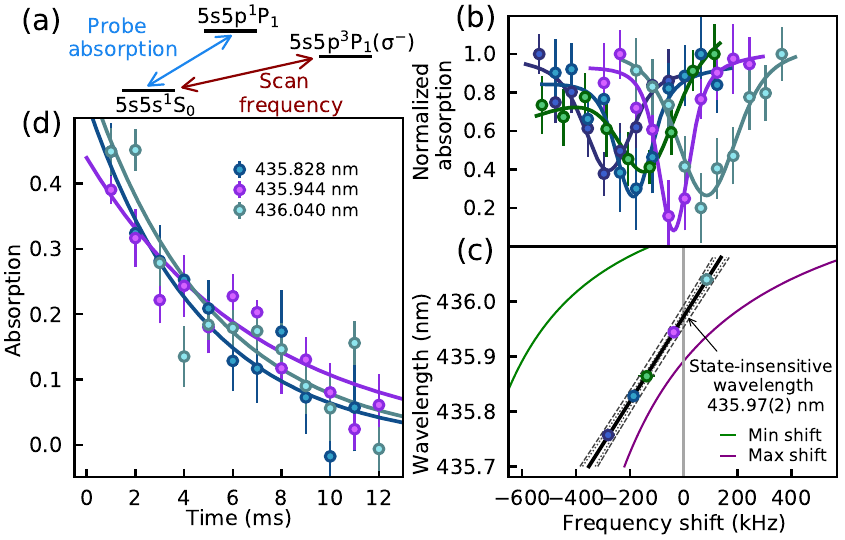}
    \caption{(a) Energy level diagram for shelving-type spectroscopy on the narrow-line intercombination transition ${^1S_0 - {}^3P_1}$ ($\sigma^{-}$). We scan the frequency of the laser driving the narrow-line transition (red arrow) and record absorption on the ${}^{1}S_{0} - {}^{1}P_{1}$ transition (blue arrow). (b) Shelving-type spectroscopy data by scanning the shelving laser frequency across the atomic transition for various wavelengths of the blue-detuned trapping field, as referenced in (c). The absorption data is normalized to the maximum value of each spectroscopy scan. We fit spectroscopy data using the Voigt model to determined the resonant frequency. (c) Measured transition frequency shift of the ${}^{1}S_{0} - {}^{3}P_{1, m=-1}$  transition as a function of the blue-detuned trapping wavelength. Fitting to a linear model, we determined the state-insensitive wavelength at the zero crossing to be 435.97(2)~nm. We include the predicted minimum(maximum) frequency shift at different blue-detuned wavelengths shown by the green(purple) line. The measured frequency shifts stay within this range (see Appendix~\ref{app:blue_magic_exp}). (d) Trap lifetime measurements at wavelengths below (435.828 nm), near (435.944 nm), and above (436.04 nm) this measured state-insensitive wavelength.}
  \label{fig:shelving_spectroscopy}
\end{figure}

Using this external beam, we implement a shelving-type spectroscopy by sending a $\sigma^{-}$-polarized pulse onto the trapped atoms, thereby shelving a fraction of them in the ${}^{3}P_{1, m=-1}$ state. Before these atoms decay back to the ground state, we measure absorption through the fiber by probing the 461-nm transition [see Fig.~\ref{fig:shelving_spectroscopy}(a)]. The case where the shelving beam is resonant to the ${}^{1}S_{0} - {}^{3}P_{1,m=-1}$ transition is indicated by a drop in the absorption measurement since the excited atoms will not absorb any of the guided probe light [see Fig.~\ref{fig:shelving_spectroscopy}(b)]. For these measurements, the 461-nm probe beam power is 50(3)~pW with a pulse duration of 15~$\mu$s, less than the excited state lifetime of 21~$\mu$s, and is sent immediately after the 4~$\mu$s shelving pulse, i.e., when the $5s5p\,{}^{3}P_{1,m=-1}$ excited state is maximally populated (see Appendix~\ref{app:shelving_calibration}).

We confirm the state-insensitive nature of the trapping fields by performing shelving-type spectroscopy at various blue-detuned wavelengths around the theoretically predicted value of the magic wavelength [see Fig.~\ref{fig:shelving_spectroscopy}(b,c)]. For reference, we apply the 200(11)~mG in the vertical direction with the MOT coils and initially measure the unperturbed resonance of the ${}^{1}S_{0}-{}^{3}P_{1,m=-1}$ transition with free-space, spin-resolved shelving-type spectroscopy using our cold atom cloud 3~ms after the MOT beams have been turned off (see Appendix~\ref{app:shelving_calibration}). We then perform shelving-type spectroscopy through the fiber with the atoms trapped using repulsive wavelengths from 435.75~nm to 436.05~nm and keeping the total power of the repulsive trapping field fixed at 1.03(1)~mW with nominally linear vertically oriented polarization at the trapping minima. Each absorption curve is fit to a Voigt profile to account for the Doppler broadened atomic cloud temperature and power broadening from the external shelving beam with an intensity of 2.3~mW/cm$^{2}$ at the position of the atoms (see Appendix~\ref{app:shelving_calibration}). We determine the AC-Stark shift of the blue-detuned trapping field by subtracting out the initial free-space reference measurement. 

At wavelengths below (above) the magic wavelength, we observe a negative (positive) frequency shift of the ${}^{1}S_{0} - {}^{3}P_{1,m=-1}$ resonance, while a state-insensitive trap for the ${}^{1}S_{0} - {}^{3}P_{1,m=-1}$ transition is realized for a wavelength of 435.97(2)~nm. In order to check if the wavelength of the blue-detuned trapping field influences the trap potential, we perform lifetime measurements for three settings of the blue-detuned wavelength [see Fig.~\ref{fig:shelving_spectroscopy}(d)]. Each measurement falls within the error bars of the lifetime in the ``lattice'' setting in Figure~\ref{fig:state_insensitive_trapping}(b), which shows that the trap is unaffected by the exact wavelength of the blue-detuned trapping field nearby the magic wavelength.

The experimental state-insensitive trapping wavelength is systematically shifted by 0.033(7)\% from the predicted magic wavelength value of 435.827(25)~nm (see Appendix~\ref{app:blue_magic_exp}). This is most likely due to the fact that the blue-detuned trapping field is not completely linearly polarized at the position of the atoms. Given the theoretical scalar, vector, and tensor polarizabilities, we propagate the experimental uncertainty in the trapping field polarizations to obtain a systematic uncertainty of the predicted magic wavelength at $436.0^{+0.2}_{-0.1}$~nm, in agreement with the measured magic wavelength of 435.97(2)~nm.

The results shown in Fig.~\ref{fig:state_insensitive_trapping} were obtained with a magic wavelength of 435.97~nm for the blue-detuned trapping field and an attractive magic wavelength of 473.251~nm for the red-detuned trapping field, thereby demonstrating state-insensitive trapping of ${}^{88}$Sr around a nanofiber. 

\section{Outlook}

Optically trapped neutral atoms in the evanescent fields of a tapered nanofiber have been studied extensively in the hopes of developing novel technologies, such as a quantum internet~\cite{Kimble2008}. Research into collective radiation~\cite{Pennetta2022}, chiral quantum optics~\cite{Petersen2014, Lodahl2017}, quantum memories of light~\cite{Patnaik2002, Gouraud2015, Sayrin2015}, and infinite range atom-atom interactions~\cite{Solano2017}, have all progressed the goal of developing revolutionary quantum technologies. This work is the first realization of alkaline-earth elements trapped along photonic devices, which opens the door for novel contributions to many of these fields. For example, the ratio of our lattice spacing to the 689-nm transition is $\approx$0.3, which is attractive for studying super- and sub-radiance effects. Furthermore, the ${}^{1}S_{0}-{}^{3}\!P_{0}$ clock transition of strontium-88 presents a platform for quantum networks with long lived memory by trapping and addressing individual strontium atoms near the surface of an optical nanofiber. There are challenges, however, to effectively use the clock transition of near-surface trapped strontium atoms for time keeping or quantum information processing due to large shifts from the van der Waals and Casimir-Polder potentials~\cite{Cook2017, Martin2017}.

Additionally, strontium's narrow ${}^{1}\!S_{0}-{}^{3}\!P_{1}$ cooling and ${}^{1}\!S_{0}-{}^{3}\!P_{0}$ clock transitions provide unique high-resolution spectroscopic tools for quantum science. For example, the van der Waals and Casimir-Polder potentials induce atomic energy shifts at various surface-to-atom distances. These shifts are on the order of 50-100~kHz and are thus detectable using the narrow, 7.6-kHz transition of $^{88}$Sr~\cite{Martin2017,Cook2017}. Building on the work in this paper, it is possible to vary the trap distance from the dielectric surface between 200~nm and 300~nm. In this case, shelving-type spectroscopy as above can be used to probe the van der Waals and Casimir-Polder effects, specifically in the cross-over region between the two. However, the small diameter of the fiber lessens the measurable surface effects. Alternative waveguide photonic structures, discussed below, afford the ability to generate trapping potentials $<$200~nm from the dielectric surface and have a larger surface area to better observe these effects spectroscopically.

We envision that two-color evanescent field traps for alkaline-earth atoms, as demonstrated here, could also be implemented in the evanescent field above nanowaveguides in photonic integrated circuits. These structures could solve any vibrational heating mechanisms which limit the trapping lifetime in nanofiber-based traps since the devices would be integrated circuits on a chip instead of a suspended fibers. Here, micro-ring resonators are particularly appealing because they may allow one to realize ring-shaped trapping potentials. If strontium atoms were loaded and cooled to quantum degeneracy in such a chip-based ring-trap, the corresponding matter wave could then be coherently split into two clouds that rotate with opposite sense~\cite{Andersson2002, Mazzoni2015}. Recombining the clouds, one would then obtain a gyroscope that is potentially orders of magnitude more sensitive than laser gyroscopes~\cite{Wu2007, Moan2019, Krzyzanowska2022}. For these applications, the small collisional scattering length of $^{88}$Sr, which is almost two orders of magnitude smaller than $^{87}$Rb, would reduce collisional dephasing significantly~\cite{Jo2007, Japha2021}. Encouragingly, ultra-high Q ring resonators that operate at the blue strontium wavelength have recently been demonstrated~\cite{Liu2018}. Moreover, photonic devices with an on-chip MOT for strontium are currently being developed~\cite{Chauhan2022}. When combined with the state-insensitive evanescent field trapping of strontium demonstrated here, this then sets the stage for future photonic-atomtronic integrated circuits with potential applications ranging from quantum sensing to complex quantum many-body physics with light.

\begin{acknowledgements}

We would like to thank P. Lauria for helpful insight, discussions, and assistance building the experimental apparatus. We are grateful to D. Steck for productive and insightful conversations. We acknowledge the support of the Office of Naval Research under Grants No. N00014-20-1-2513 and N00014-20-1-2693 and NSF Grant PHY-2012068.
This research was supported in part through the use of University of Delaware HPC Caviness and DARWIN computing systems: DARWIN - A Resource for Computational and Data-intensive Research at the University of Delaware and in the Delaware Region, Rudolf Eigenmann, Benjamin E. Bagozzi, Arthi Jayaraman, William Totten, and Cathy H. Wu, University of Delaware, 2021~\cite{udel}. We acknowledge funding by the Alexander von Humboldt Foundation in the framework of the Alexander von Humboldt Professorship endowed by the Federal Ministry of Education and Research.

K.T. and G.K. contributed equally to this work.

\end{acknowledgements}

\begin{appendix}

\section{Evanescent Fields}
\subsection{\label{app:trap_potentials}Nanofiber-based trapping potentials}

\begin{figure}[t!]
    \includegraphics{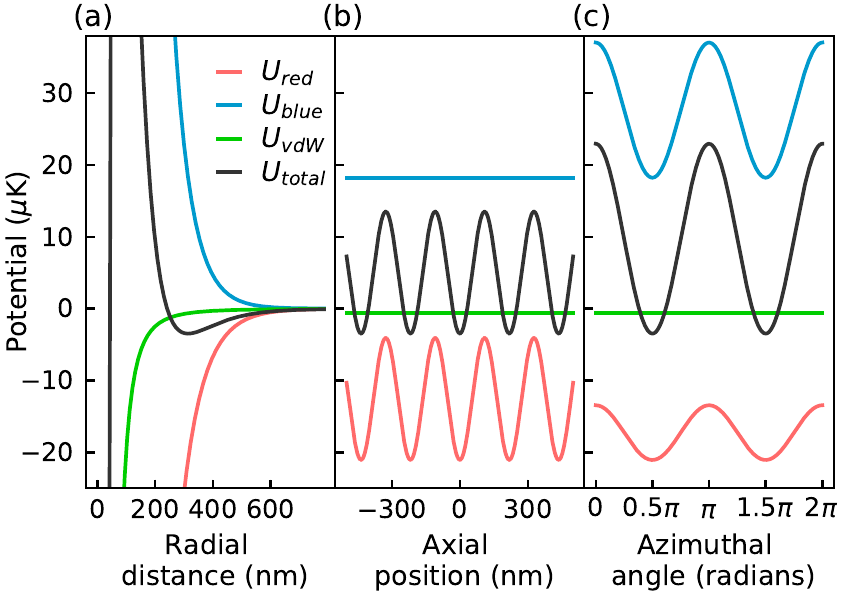}
    \caption{(a) Computed total trapping potential of the nanofiber-based trap (black curve) as a function of the radial distance from the nanofiber surface for our experimental parameters. The total potential arises from the sum of the red-detuned and blue-detuned optical potentials (red and blue), as well as the van der Waals potential (green). At the radial distance of the total potential minimum, we show the trapping potentials in (b) the axial direction along the nanofiber and (c) the azimuthal direction around the nanofiber.}
    \label{fig:all_potentials}
\end{figure}

A complete optical dipole trap around the nanofiber confines the atoms in all radial, axial, and azimuthal directions~\cite{Vetsch2010a}. As discussed in the main text, we achieved this by employing a red-detuned attractive light field combined together with a blue-detuned repulsive light field. Due to the smaller decay length of the blue-detuned evanescent field compared to the red-detuned field, the resulting total potential has a minimum at some distance from the nanofiber surface, which provides the radial confinement. In this radial direction, we have to take into consideration the van der Waals potential of a strontium atom near the nanofiber dielectric surface. For a flat silica wall, an approximation for our configuration, this potential is given by 
$$U_{\rm{vdW}}(r) = \frac{\epsilon - 1}{\epsilon + 1} \frac{C_{3}}{r^3},$$
where $\epsilon$ is the dielectric permittivity and $C_{3}$ is the van der Waals coefficient for atom-surface interaction~\cite{Burke2002}, and we consider the value of $C_{3}$ for strontium from reference~\cite{Derevianko2010}. We then scale the van der Waals potential down appropriately to account for the fact that the nanofiber has a cylindrical shape~\cite{Kien2004a,Salem2010}. Taking all contributions into account, the trapping potential in the radial direction has a depth of about $3.5$~$\mu$K at 320~nm from the nanofiber surface [see Fig.~\ref{fig:all_potentials}(a)].

Along the nanofiber axis, the axial confinement is achieved by an optical lattice formed by two counter-propagating red-detuned beams calibrated to have the same power at the nanofiber waist. In addition, the blue-detuned running wave provides a constant repulsive potential, thereby, together with the red-detuned beams, forming separate regularly-spaced trapping sites along the nanofiber [see Fig.~\ref{fig:all_potentials}(b)].

Azimuthal confinement of the nanofiber-based trap is generated by setting orthogonal quasi-linear polarizations for the red-detuned and blue-detuned trapping fields. Around the nanofiber, with quasi-linear polarization, the evanescent field intensity is stronger along the polarization direction. Therefore, another benefit of setting the polarization of the trapping fields to be orthogonal is to achieve a deeper trapping potential with less optical power because the azimuthal intensity maximum of the red-detuned field overlaps with the azimuthal intensity minimum of the blue-detuned field [see Fig.~\ref{fig:all_potentials}(c)]. Furthermore, the longitudinal component of the electric field vanishes at the trap locations that lie orthogonal to the quasi-linear polarization axis. This ensures the blue detuned optical field is linearly polarized at the trap locations~\cite{Kien2004}. For the red-detuned trapping field, the longitudinal electric-field components of the counter-propagating beams interfere destructively at the trap locations, resulting in purely linear polarization in the resulting optical lattice sites.

\subsection{\label{app:pol_calibration}Polarization calibration}

\begin{figure}[t!]
    \includegraphics[width=\columnwidth]{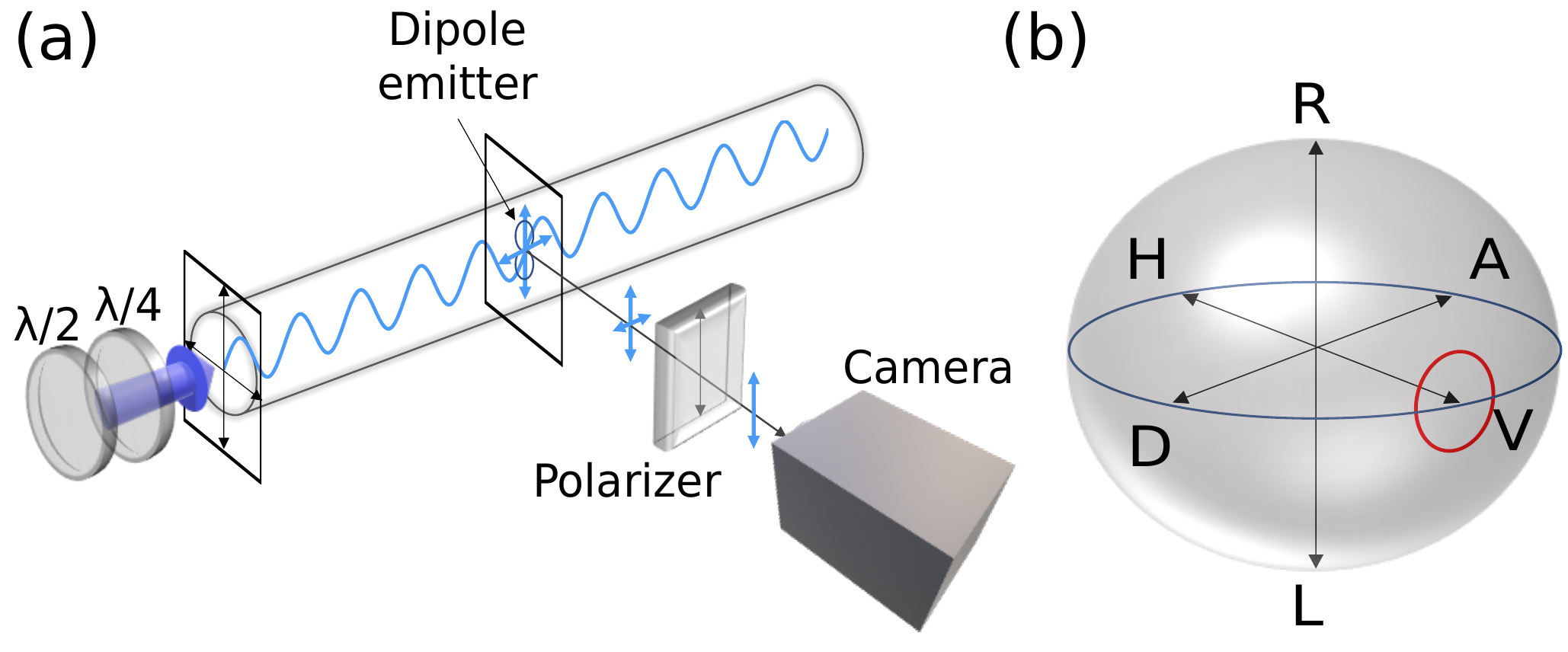}
    \caption{(a) Experimental setup for polarization calibration of the trapping beams. We adjust the input polarization by rotating the half-wave ($\lambda/2$) and quarter-wave ($\lambda/4$) plates. A camera is used to record the fluorescence scattering from the fiber through a polarizer, whose axis is set perpendicular to the fiber's axis. (b) Uncertainty in the polarization state of light in the nanotapered fiber (red circle) depicted on the Poincar\'e sphere. The radius of the circle is inversely proportional to the visibility of the polarization measurement.} 
    \label{fig:pol_setup}
\end{figure}

To calibrate the polarization of the propagating fields in the nanofiber, we rely on imaging of the Rayleigh scattering of the dipole emitters in the fiber material \cite{Goban2012a, Vetsch2010a, Joos2019} and across the surface. Assuming a polarization maintaining scattering process, the dipole excitations are directly related to the polarization of the guided fields. A camera images the side of the fiber encoding the polarization angle $\theta$ into an intensity measurement that is $\propto \sin^{2}(\theta)$, the dipole radiation pattern. At the side camera, there is a maximum (minimum) intensity when the dipole is excited horizontally (vertically) [see Fig.~\ref{fig:pol_setup}(a)].

Due to a longitudinal electric field component along the fiber axis, it is possible to excite the dipoles in this direction~\cite{Kien2004}. We place a linear polarizer aligned orthogonal to the fiber axis (vertically) to filter out this non-zero component [see Fig.~\ref{fig:pol_setup}(a)]. We use a half-wave plate and quarter-wave plate to adjust the polarization of the fiber guided fields. For all three trapping fields and probe field, we first incrementally adjust the half- and quarter- wave plates to achieve a minimum intensity $I_{\rm{min}}$. Then, we incrementally adjust the waveplates again to reach a maximum intensity $I_{\rm{max}}$. From these measurements, we calculate the visibility of the 436-nm beam, as well as the forward propagating (+) and the backward propagating (-) 473-nm beams as $V=(I_{\rm{max}} + I_{\rm{min}})/(I_{\rm{max}} - I_{\rm{min}})$, to be \{$V_{436}, V_{473+}, V_{473-}$\} = \{70, 61, 45\}~\%.

The visibility defines an uncertainty from purely horizontal or vertical polarization. The maximal intensity occurs when the polarization angle, relative to vertical, is $\theta = \pi/2$ while the minimum intensity should occur at $\theta = 0$, assuming a dipole radiation pattern. However, an error in the polarization means the minimum intensity occurs at $\theta_{\rm{err}}$ instead

\begin{align*}
    V &= \frac{1 + \sin^{2}(\theta_{\rm{err}})}{1 - \sin^{2}(\theta_{\rm{err}})}\nonumber\\
    \theta_{\rm{err}} &= \arcsin{\sqrt{\frac{1-V}{1+V}}}.\nonumber
\end{align*}

The $\{\theta_{\rm{err,436}},\theta_{\rm{err,473+}},\theta_{\rm{err,473-}}\}=\{24.8,29.5,38.0\}^{\circ}$ measured in our setup defines an angle of deviation on the Poincar\'e sphere from purely vertical or horizontal polarization [see Fig.~\ref{fig:pol_setup}(b)]. Unfortunately, circularly polarized and diagonally polarized propagating fields will induce a similar dipole oscillation with phase differences undetectable in our single camera setup~\cite{Tkachenko2019, Joos2019}. Being unable to discern the two, a circle can be drawn on the Poincar\'e sphere around purely horizontal and vertical polarization for the possible polarization of the propagating field. The radius of this circle is defined by $\theta_{\rm{err}}$.

The uncertainty in polarization of the trapping fields affects the trap parameters as well as the magic wavelength measurement. We propagate these uncertainties for the trap parameters reported in the main text. In Appendix \ref{app:blue_magic}, we discuss the systematic shift in the experimental magic wavelength.

\subsection{\label{app:power_calibration}Power calibration}

For the absorption measurements and trapping potentials, we require precise knowledge of the trapping and probe field powers at the waist of the nanofiber. Due to our vacuum chamber geometry, the fiber tails on either side of the waist incur different bending losses. We begin by denoting three sections of the fiber as $A$, $B$, and $C$ where $A$ is one of the tails, $B$ is the nanofiber waist, and $C$ is the other fiber tail. We send two counter propagating reference beams from $A\rightarrow C$ and $C\rightarrow A$ through the entire fiber and measure each power at the corresponding output $P_{A\rightarrow C}$ and $P_{C\rightarrow A}$, respectively. The power of each beam is then set so the fluorescent scattering at the fiber waist is equal, $P_{A\rightarrow C}(B) = P_{C\rightarrow A}(B)$. With the percentage of power lost in each section denoted as $L_{A}$, $L_{B}$, and $L_{C}$, we then have a relationship between the losses in each tail given as
$$\frac{P_{A\rightarrow C}(C)}{1 - L_{C}} = \frac{P_{C\rightarrow A}(A)}{1 - L_{A}}.$$
It should be noted that the loss $L_{A}$ includes all losses up to and including the adiabatic taper to the waist while $L_{B}$ only considers losses across the $\approx$1~mm region that overlaps with our MOT cloud and is thus negligible since $L_{B}\ll L_{A}, L_{C}$.

This provides us with a relationship between $L_{A}$ and $L_{C}$. At this point, we also need a measurement of the total loss through the fiber, which requires knowledge of the coupling efficiency from our free-space beam. We characterize the fiber coupling using a straight, $\approx$0.5~m identical fiber so that the propagation losses are negligible. We repeatedly perform the fiber end splicing, placement, and coupling procedure that identically to that of the nanofiber. Through this process, we find a repeatable coupling efficiency of $(1-L_{\rm{coupling}}) = 80(5)\%$ into the optical fiber. We then measure $P_{\rm{in}}$ before coupling into the fiber and $P_{\rm{out}}$ at the output of the fiber to determine the full relationship as $P_{\rm{out}} = P_{\rm{in}}(1-L_{\rm{coupling}})(1-L_{A})(1-L_{C})$.

The empirical values are $P_{C\rightarrow A}(A) = 75(4)~\mu$W, $P_{A\rightarrow C}(C)=53(3)~\mu$W, and $P_{\rm{out}} = 0.53(4)\times P_{\rm{in}}$ for both directions. Using the fiber coupling efficiency of 80(5)\%, we determine the losses to be $L_{A}=32(2)\%$ and $L_{C}=3.18(12)\%$. Ultimately, we are concerned with the total power in the $\approx$1~mm region of the nanofiber waist where the MOT cloud will be overlapped. For the blue-detuned trapping field, the probe field, and the forward propagating red-detuned trapping field, this is given by $P_{A\rightarrow C}(B) = P_{\rm{out}}/(1-L_{C})$, while the counter-propagating red-detuned trapping field power at the nanofiber waist is given by $P_{C\rightarrow A}(B) = P_{\rm{out}}/(1-L_{A})$.

\section{\label{app:heating_rates}Heating rates}

Heating of the atoms in our nanofiber trap has contributions from various sources. The trapping light, being nearby the ground state of the 461-nm resonance and the excited states of the 436-nm and 473-nm resonances, results in spontaneous emission in both states. The spontaneous emission heating rate from a photon with recoil energy $E_{r}=(\hbar k)^{2}/2m$ is given by $E_{r}\Gamma_{\rm{sc}}/k_{B}$,
and the scattering rate $\Gamma_{\rm{sc}}$ is
$$\Gamma_{\rm{sc}}=U(r_{\rm{atoms}})\left(\frac{\Gamma}{\hbar\Delta}\right),$$
where $U(r_{\rm{atoms}})$ is the optical potential for each individual trapping field, $\Delta$ is the detuning from the nearest transition, and $\Gamma$ is the natural linewidth of the transition. Since our trapping wavelengths are close in proximity to both ground and excited state transitions, we consider the heating rates for all relevant detunings in Table~\ref{tab:spont_emis_heating_rates}.

Because the trap position is dependent on the intensity of light at the nanofiber waist, fluctuations of the total power will cause heating in the radial direction. This heating is dominated at the radial trap frequency and scales as $\nu_{r}^{2}$~\cite{Grimm2000}. We implement intensity stabilization for each of our trapping beams prior to coupling light into our nanofiber, which stabilizes the input beam powers up to 100~kHz. The main source of intensity fluctuations are on the order of hours due to coupling drifts into the fiber tails, and thus, have no effect on heating rates during the experiment. With the input intensities stabilized, we estimate a heating rate of $\approx5.3\times10^{-4}$~$\mu$K/ms for the 436-nm trapping field and $\approx2.5\times10^{-4}$~$\mu$K/ms for the 473-nm trapping fields. Neither of these rates limits the trapping lifetime.

In addition to intensity fluctuations, the optical lattice generated from the counter-propagating red-detuned trapping lasers contributes heating proportional to the large axial trap frequency $\propto\nu_{\rm{z}}^{4}$. The measured phase noise of $10^{-10}$~rad$^{2}$/Hz at the axial trap frequency contributes $6\times10^{-1}$~$\mu$K/ms, about three times higher than the largest spontaneous emission heating rates. However, the potential well in the axial direction is $\approx$20~$\mu$K in depth and will not limit our trapping lifetime unless all of the axial phase noise couples into the radial trapping potential where the potential depth is not as low, which is highly unlikely. This heating source can be reduced with interferometric stabilization.

There is also a known issue with heating from vibrational and torsional modes of the suspended fiber structure~\cite{Hummer2019}. For our experimental parameters this yields a heating rate of $\approx$16~$\mu$K/ms, which limits the lifetime of our trap to $<$1~ms. Unfortunately, these calculations and the phonon-photon coupling parameters were calculated under the assumption that a large number of motional states were trapped. While in our experiment, given the $\approx$1~$\mu$K atom temperature and the shallow trap depth, we are trapping far fewer motional states. However, we suspect the vibrational noise is limiting our trap lifetime, especially considering this was the limiting case in the first nanofiber-based traps as well and was solved with continuously cooling the trapped atoms.

\begin{table}[t!]
\caption{\label{tab:spont_emis_heating_rates} Spontaneous emission contributions to the heating rates of the ground and excited states in the trapping light at 435.97~nm and 473.251~nm. Calculations only consider contributions to the nearest transition given in the second column.}
{\renewcommand{\arraystretch}{1.2}
\begin{tabular}{p{0.0cm}>{\centering}p{1.0cm}>{\centering}p{1.7cm}>{\centering}p{1.7cm}>{\centering}p{1.7cm}>{\centering}p{1.7cm}>{\arraybackslash}p{0.0cm}}
 & State & Resonance (nm) & Wavelength (nm) & Scattering Rate (Hz) & Heating Rate ($\mu$K/ms) &\\
\hline
& $5s5p\;^{3}\!P_{1}$ & 472.359 & 473.251 & 3.8 & $1.9\times10^{-3}$ &  \\\hline
& &  & 435.97 & 0.4 & $2.3\times10^{-4}$ & \\\hline
& & 436.293 & 473.251 & 0.02 & $9.8\times10^{-6}$ &\\\hline
& &  & 435.97 & 10.4 & $6.0\times10^{-3}$ &\\\hline\hline
& $5s^{2}\;^{1}\!S_{0}$ & 460.86 & 473.251 & 19.6 & $9.6\times10^{-3}$ &\\\hline
& &  & 435.97 & 51 & $2.9\times10^{-2}$ &\\
\hline\hline
\end{tabular}}
\end{table}

\section{Effect of the blue-detuned field}

\subsection{\label{app:blue_detuned_density_enhancement}Atom density enhancement}

\begin{figure}[t!]
    \includegraphics{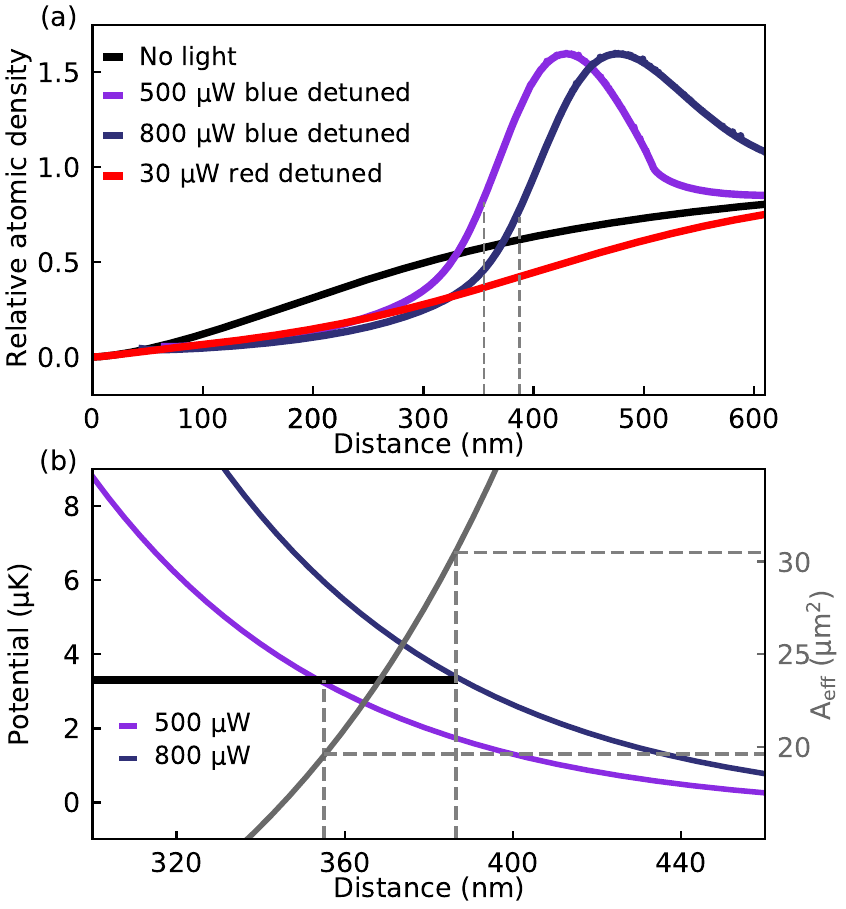}
    \caption{(a) Relative atom number density of the atoms in a MOT with 1~$\mu$K temperature as a function of distance from the nanofiber's surface in the present of different evanescent field potentials. (b) Calculated $A_{\rm{eff}}$ based on parameters experimentally determined as shown in Fig.~\ref{fig:redmot_blue_data}(c), as a function of atom-surface distance, and total radial potential with 500~$\mu$W and 800~$\mu$W of repulsive blue-detuned trapping field (purple and blue). Interestingly, the distances at which we probe the atoms coincides with the position where the total potentials are both $\approx$3.3~$\mu$K (black).} 
    \label{fig:blue_density}
\end{figure}

The number density of an atomic ensemble around the nanofiber is modified relative to the free-space number density. It has been showed that the atomic number density is lowered due to the van der Waals and light-induced potentials~\cite{Sague2007}. An attractive potential caused by the van der Waals and a red-detuned beam significantly decreases the number density around the nanofiber. Interestingly, in a repulsive blue-detuned potential, the number density is lowered at short distances, but it is greatly enhanced at further distances. We calculate the relative number density of an atomic ensemble in the presence of a potential $U(r)$ with~\cite{Kien2009}

$$\frac{n(r)}{n_{0}} = \frac{1}{k_{B}T}\int^{\infty}_{0}e^{-E/k_{B}T}\frac{\sqrt{E}}{\sqrt{E-U(r)}}dE,$$
where $n_{0}$ is the free-space optical depth as $r\rightarrow\infty$, $T$ is the atomic cloud temperature, and $U(r)$ is the radially dependent potential. In Fig.~\ref{fig:blue_density}(a), we solve this integral numerically for various potentials $U(r)$, for an atomic cloud temperature of $1~\mu$K.

The black line shows the atomic density assuming only $U_{\rm{vdW}}(r)$, which decreases the atomic density close to the fiber. When also taking the potential generated by a red-detuned trapping field with a power of 30~$\mu$W into account $U_{\rm{vdW}}(r) + U_{\rm{red}}(r)$, we obtain the density distribution shown as solid red line. Further considering 500 (800)~$\mu$W of blue-detuned trapping field $U_{\rm{vdW}}(r) + U_{\rm{blue}}(r)$, the repulsive force increases the atomic density some distance from the fiber surface as indicated on the purple (blue) line.

With the atom cloud around the nanofiber, the fraction of the guided probe beam that is absorbed depends on the decay of the evanescent field from the surface, the density of the atom cloud, and the shift of the atomic resonance induced by the van der Waals potential. Closer to the fiber, the atomic density is lower, but the optical depth per atom ($OD_{\rm{atom}}$) is higher. Inversely, farther from the fiber, the atomic density is higher, but the $OD_{\rm{atom}}$ is lower. The contribution from the van der Waals potential is only appreciable below distances of $\approx$50-100~nm. Balancing all of these contributions, there exists an average distance from the surface where we are probing the atoms.

\begin{figure}[t!]
    \includegraphics{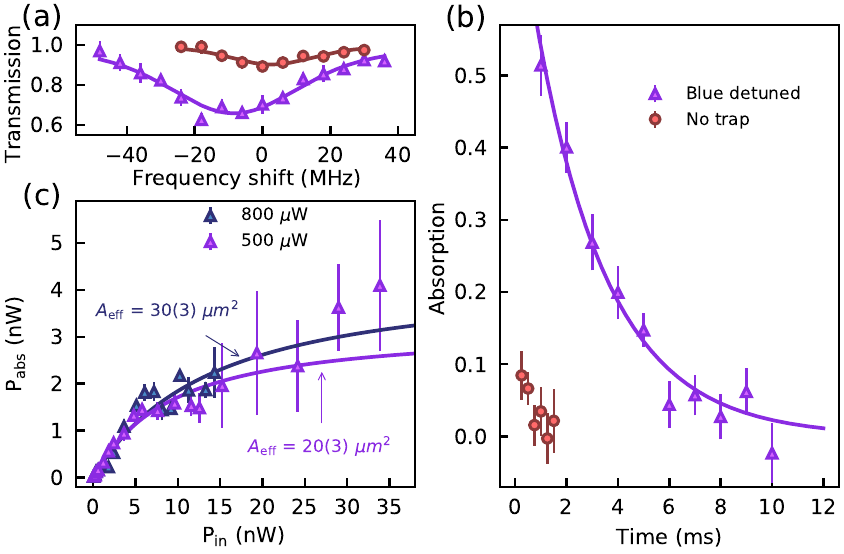}
    \caption{(a) Spectroscopy of the $5s^2$ $^1S_0$ - $5s5p$ $^1P_1$ transition of a cloud with a temperature $\approx$1 $\mu$K with (purple) and without (red) 500~$\mu$W of 436-nm blue-detuned evanescent trapping field. (b) The absorption decays with a time constant of 2.8(2) ms in the presence of the blue-detuned field. (c) Saturation absorption measurement at different powers of 436-nm light. We fit the data to the generalized Beer's law for absorption (see Appendix~\ref{app:evanescent_fields_number}).
    Each fit yields an ensemble-averaged effective mode area $A_{\rm{eff}}$, from which we can deduce an effective distance between the atoms and the nanofiber surface.}
    \label{fig:redmot_blue_data}
\end{figure}

With only the blue-detuned evanescent field, there is no potential minimum and saturation absorption measurements can be performed without concern for removing atoms from the trap. From the measurements in Fig.~\ref{fig:redmot_blue_data}(c), we extrapolate the effective mode area $A_{\rm{eff}}$ of the evanescent probe beam through the fiber (see Appendix \ref{app:evanescent_fields_number}). This provides us with a first order approximation to the average distance where the atoms are being probed. With 500~$\mu$W (800~$\mu$W) of blue-detuned field, we measure this to be $A_{\rm{eff}}=20(3)~\mu\rm{m}^{2}$ ($30(3)~\mu\rm{m}^{2}$) corresponding to distances of 355(11)~nm (386(9)~nm) from the fiber surface [see Fig.~\ref{fig:blue_density}(b)]. Interestingly, the blue-detuned optical potential in each case is $\approx$+3.3~$\mu$K and the atomic density is $\approx$0.7, relative to 0~$\mu$K potential and an atomic density of 1 in the free falling atomic cloud. This implies that, without a trapping potential, atoms from our cold atomic cloud are stopped by the potential barrier in such a way that we consistently probe them where the potential barrier is $\approx$3.3~$\mu$K. With larger blue-detuned intensities, the distance from the fiber surface where the barrier is $\approx$3.3~$\mu$K will increase. Therefore, the atoms will be exposed to the same optical intensity, despite increasing the total blue-detuned power.

\subsection{\label{app:blue_detuned_light_shift}$^{1}\!S_{0}-^{1}\!P_{1}$ Transition Spectroscopy}

With only the blue-detuned, repulsive field in the fiber at powers greater than $500$~$\mu$W, we observe $\approx$-8~MHz shift in the ${}^{1}S_{0}-{}^{1}P_{1}$ resonance [see Fig.~\ref{fig:redmot_blue_data}(a)]. We verify that blue-detuned field is indeed the cause of the shift by varying the power from 0 to 500~$\mu$W and performing spectroscopy measurements through the fiber. 

\begin{figure}[!t]
    \includegraphics{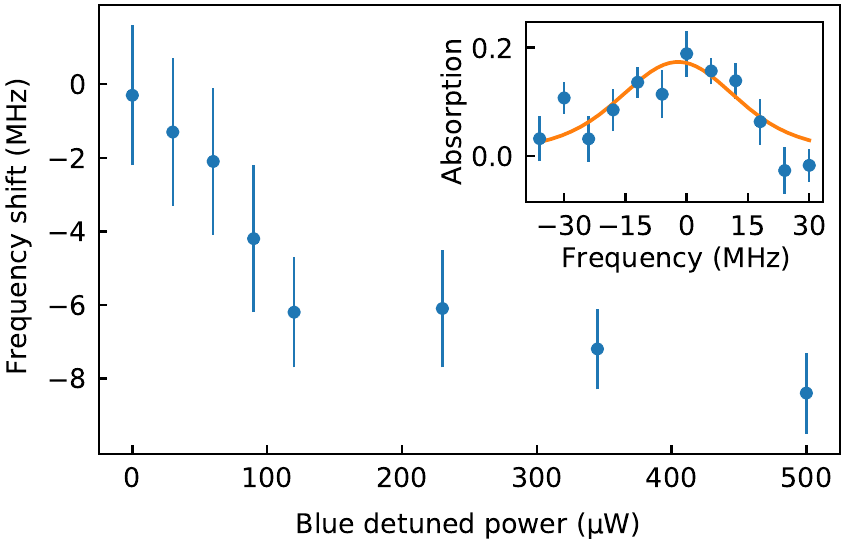}
    \caption{Measurements of the frequency shift of the probing transition as a function of the blue-detuned field power. The inset shows the absorption spectrum where the blue-detuned power is 60~$\mu$W.
    } 
    \label{fig:blue_shift}
\end{figure}

Additionally, the 461-nm spectroscopy data does not exhibit any broadening of the natural linewidth of $\Gamma = 2\pi\times30.5$~MHz. This may seem counter-intuitive since the shallow trap depth allows for uncertainty in the atom positions in the trap, thus exposing the atoms to different intensities of the blue-detuned field. In principle, this induces different shifts for each trapped atom effectively broadening the linewidth, however, the intensity difference across the position uncertainty likely induces a small enough shift relative to the natural linewidth to not be detected. Calculating these shifts is difficult for the excited $^{1}\!P_{1}$ state due to reasons discussed in the main text.

\section{\label{app:evanescent_fields_number}Atom Number Estimation}

\begin{figure}[t!]
    \includegraphics{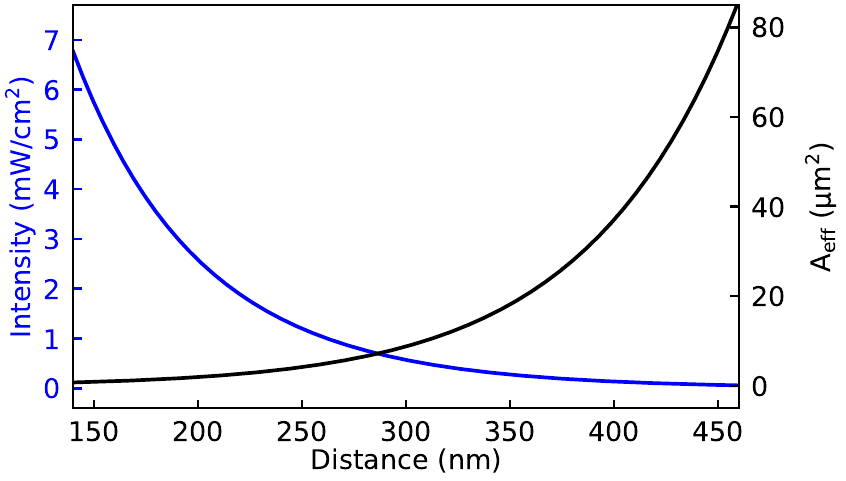}
    \caption{Intensity of a 50~pW probe beam at 461~nm as a function of distance from the fiber surface (blue curve). For reference, the saturation intensity of the probing transition is $\approx$40.7~mW/cm$^2$. Independent of the light power, the $A_{\rm{eff}}$ is a function of distance (black curve). Atom-surface separation can be extracted from the $A_{\rm{eff}}$, which is obtained from the saturation absorption measurement. }
    \label{fig:Intensity_area}
\end{figure}

For a constant total power $P$ at the waist of the nanofiber, the evanescent fields have an intensity given by $I(r) = P/A_{\rm{eff}}(r)$, thus the intensity and effective mode area $A_{\rm{eff}}$ are inversely proportional to each other (see Fig.~\ref{fig:Intensity_area}). A probe beam resonant with the atomic transition will have an intensity given by $I(r_{\rm{atoms}}) = P/A_{\rm{eff}}(r_{\rm{atoms}})$, where $r_{\rm{atoms}}$ is the distance of the atoms from the fiber surface. Furthermore, the atoms have an optical depth $OD_{\rm{atom}} = \sigma_{0}/A_{\rm{eff}}(r_{\rm{atoms}})$, where $\sigma_{0} = 3\lambda^{2}/2\pi$ is the on-resonant scattering cross section. 

This is also the cooperativity of a single atom with the evanescent probe field. For the 461~nm probe field and the fiber diameter of 230~nm, we see the single atom cooperativity approach $C_{1}=1$ around $r=30$~nm. This close to the fiber surface, however, the van der Waal's potential ($\propto 1/r^{3}$) dominates over the repulsive field ($\propto 1/r^{2}$) and a trapping potential cannot be formed. One benefit of using the waveguide confined modes is that $C_{1}$ extends across the length of the fiber allowing $N$ atoms to interact with the probe field so the total cooperativity becomes $N\times C_{1}$. In practice, we can trap further from the fiber surface to minimize the effect of the surface potentials, while sacrificing the single atom cooperativity. One can then counteract the loss of cooperativity by increasing the number of atoms that are trapped along the waveguide.

\begin{figure}[!t]
    \includegraphics{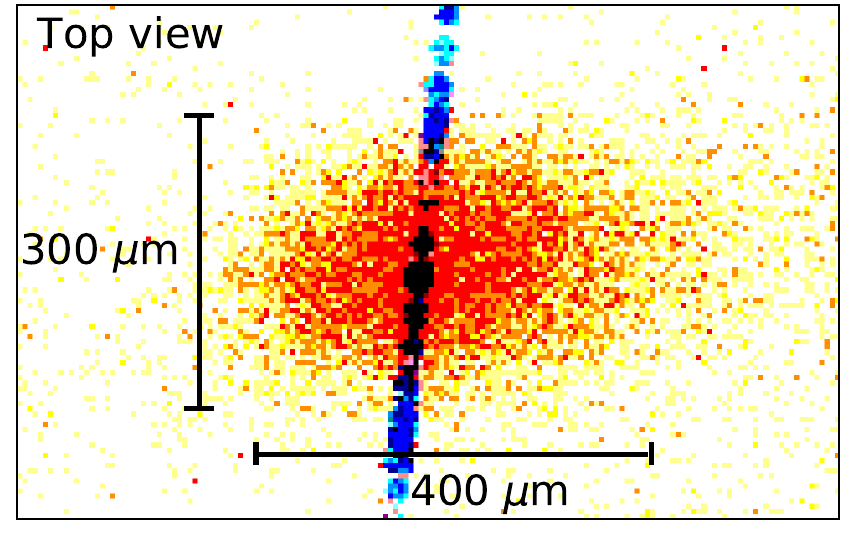}
    \caption{ Top view of the cold atomic cloud's $\approx$461~nm fluorescence (red) around the nanofiber scattering (blue trace). These cold atoms explore a trapping region of $(x, y, z) = 500~\rm{nm}\times325~\rm{nm}\times230~\rm{nm}$ for a single lattice site and there are $\approx$3000 sites, given by the size of cold atomic cloud from the MOT. The densest part of the cold atomic cloud gives enough fluorescence signal to be detected by our imaging setup and leads to an apparent transverse size of the MOT cloud of about 300~$\mu$m, in contrast to that more accurate from absorption imaging of about 600~$\mu$m, as shown in Fig.~\ref{fig:state_insensitive_trapping}(d).}
    \label{fig:atom_fiber_pic}
\end{figure}

With atoms spread along the fiber axis, the probe beam will decrease in total power as each atom absorbs a fraction of the propagating field (see Fig.~\ref{fig:atom_fiber_pic})~\cite{Vetsch2010a}. For a total number of atoms $N$, assuming linear density along the fiber of length $L$, this power decrease along the fiber axis $z$, is given by~\cite{Goban2012a}

$$\frac{dP}{dz} = -\frac{N}{L}\frac{\sigma_{0}}{A_{\rm{eff}}(r_{\rm{atoms}})}\frac{P(z)}{1+P(z)/P_{\rm{sat}}}.$$

\begin{figure}[!t]
    \includegraphics{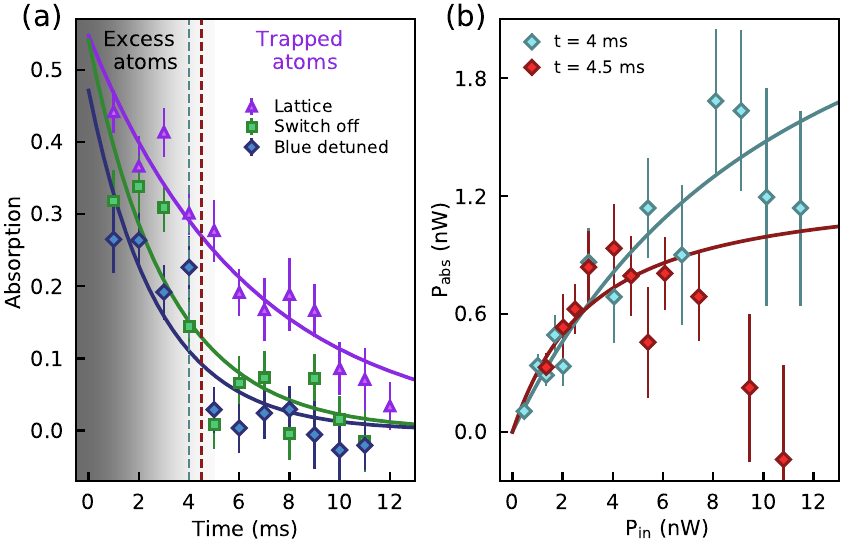}
    \caption{(a) Atomic-absorption lifetime measurements. In the \emph{switch-off} configuration the lattice beams are turned off for 200~$\mu$s shortly before probing atomic absorption. As in Fig.~\ref{fig:state_insensitive_trapping}, we compare lifetimes in three settings: switch-off, lattice trap and blue-detuned field only. (b) Saturation absorption measurements at holding times of ${t = 4}$~ms and ${t = 4.5}$~ms. Strontium atoms strongly absorb and scatter photons of the high probing intensities inherent to these measurements. Thus these absorption data are a result of atomic absorption enhancement by the blue-detuned field. By $\approx$4.5~ms and higher probing powers, the atoms scatter faster than they absorb incoming light.}
    \label{fig:trap_sat_abs}
\end{figure}

Solving this differential equation with boundary conditions of $P(0) = P_{\rm{in}}$ and $P(L) = P_{\rm{out}}$ and letting $P_{\rm{abs}} = P_{\rm{in}}-P_{\rm{out}}$ results in a LambertW (ProductLog) function. Through a saturation absorption measurement, where the input power is varied above a saturation threshold of all the trapped atoms, the LambertW function can be fit for the total number of atoms, $N$, the effective mode area at the atoms $A_{\rm{eff}}(r_{\rm{atoms}})$, and the saturation power $P_{\rm{sat}}$. At large powers $P_{\rm{in}}\gg P_{\rm{sat}}$, if all the atoms are saturated, a fixed amount of the total power will be absorbed resulting in a plateau of the absorption measurements. Unfortunately, strontium is too light of an element to remain in the trap when probing with saturation-level powers. Performing these measurements at various holding times, we confirm that excess atoms from the atomic cloud are refilling the trap at up to $\approx$4~ms (see Fig.~\ref{fig:trap_sat_abs}).

As an alternative to saturation absorption measurements, which are difficult to perform reliably given strontium's lightness compared to cesium, we estimate the number of trapped atoms using the optical depth. From the spectroscopy data in Fig.~\ref{fig:state_insensitive_trapping}(c), we fit to a Lorentzian spectrum parameterized with the total optical depth OD, given as
$$T(\omega) = \exp \left(-OD/ \left[ 1 + 4(\omega - \omega_{0})^{2})/\Gamma^{2} \right] \right),$$
where $\Gamma$ is the spectral linewidth~\cite{Vetsch2010a}. 

Using the estimated trapping distance of 320(10)~nm, the total OD of all trapped atoms is $N\sigma_{0}/A_{\rm{eff}}(320~\rm{nm}) = 0.25(3)$. With $A_{\rm{eff}}(320~\rm{nm})=10.5~\mu$m$^2$, we measure the number of atoms in the trap at ${t = 6}$~ms to be 27(3) atoms. Given the number of atoms in the trap falls off over time as $N_{0}e^{-t/\tau}$, we use the experimentally measured decay constant of the trap $\tau=7.4(7)$~ms to extrapolate the trapped number of atoms at 0~ms to be $N_{0}= 27/e^{-6/7.4} = 61(7)$ atoms.

\section{\label{app:blue_magic} Magic wavelength of the $5s^{2}\;^{1}\!S_{0} - 5s5p\;^{3}\!P_{1}$ transition}

\subsection{\label{app:blue_magic_theory}Theoretical calculation of the magic wavelength}

Polarizabilities can be computed using the sum-over formulas \cite{mitroy10}:
\begin{eqnarray}
    \alpha_{0}^v(\omega)&=&\frac{2}{3(2J+1)}\sum_k\frac{{\left\langle k\left\|D\right\|v\right\rangle}^2(E_k-E_v)}{     (E_k-E_v)^2-\omega^2}, \label{eq-1} \nonumber \\
    \alpha_{2}^v(\omega)&=&-4C\sum_k(-1)^{J+j_k+1}
            \left\{
                    \begin{array}{ccc}
                    J & 1 & j_k \\
                    1 & J & 2 \\
                    \end{array}
            \right\} \nonumber \\
      & &\times \frac{{\left\langle
            k\left\|D\right\|v\right\rangle}^2(E_k-E_v)}{
            (E_k-E_v)^2-\omega^2} \label{eq-pol},
\end{eqnarray}
where a quantity in $\{ \}$ brackets is a Wigner 6-j symbol and $j_k$ as the total angular momentum of the state $k$.
          The value of $C$ is given by
\begin{equation}
            C =
                \left(\frac{5J(2J-1)}{6(J+1)(2J+1)(2J+3)}\right)^{1/2}. \nonumber
\end{equation}
However, the sums over $k$ in the above formulas must include all possible transitions. While these sums converge rather rapidly in the case of dynamic polarizability, truncating these sums for highly excited states leads to a significant loss of accuracy. Therefore, we use a hybrid approach. First, we use a combination of the configuration interaction (CI) and the linearized coupled-cluster method (refereed to as the CI + all order method) to solve the inhomogeneous equation of perturbation theory in the
valence space, which is approximated as \cite{kozlov99a}

\begin{equation}
(E_v - H_{\textrm{eff}})|\Psi(v,m^{\prime})\rangle = D_{\mathrm{eff},q} |\Psi_0(v,J,m)\rangle.
\label{eq1}
\end{equation}
\cite{2009,Kestler2022}, where $v$ is the state of interest,  with the total angular momentum $J$ and projection
$m$.
$D_{\textrm{eff}}$ is the effective dipole operator
that includes RPA corrections, and $H_{\textrm{eff}}$ is the effective Hamiltonian used in configuration interaction (CI) calculations that includes all-order corrections calculated using the
linearized coupled-cluster method with single and double excitations.
The resulting parts of the wave function $\Psi(v,m^{\prime})$
with angular momenta of $J^{\prime} = J, J \pm 1$ are used to determine
the scalar and tensor polarizabilities. 

The advantage of this approach is that 
it automatically includes contributions from all possible states. However, while CI+all-order computation produces accurate energies, even small inaccuracies in theoretical energies may significantly contribute to uncertainties in the resonant terms in Eq.~(\ref{eq-pol}) close to the  magic wavelength. Therefore, we separately computed several most important terms in Eq.~(\ref{eq-pol}) 
 using  \textit{ab initio} energies and matrix elements that exactly correspond to our calculations using the inhomogeneous equation (\ref{eq1}) and determined the remaining contribution of all other states.
 Then we computed the same terms using the experimental energies \cite{NIST} and the recommended values of the matrix elements compiled in ~\cite{Kestler2022}, retaining the theoretical values of the  matrix elements where recommended values are not available.

\begin{table} [t!]
\caption{\label{tab2t} Contributions to the $5s5p\,^3\!P_1$ polarizability in atomic units at the 435.827~nm $^1\!S_0 - ^3\!P_1$ magic wavelength.
Transition energies $\Delta E$ in cm$^{-1}$ \cite{NIST} and the recommended values of the electric-dipole matrix elements $D$ (see the text)
in atomic units are also listed. $5s5p\,^3\!P_1$ $|m|=1$ polarizabilities are obtained as $\alpha(|m|=1)=\alpha_0+\alpha_2$.}
\begin{ruledtabular}
\begin{tabular}{lccccc}
\multicolumn{1}{c}{Contrib.}& \multicolumn{1}{c}{$\Delta E$}&  \multicolumn{1}{c}{$D$}& \multicolumn{1}{c}{$\alpha_0$}&  \multicolumn{1}{c}{$\alpha_2$} &\multicolumn{1}{c}{$\alpha(|m|=1)$}  \\
  \hline
$5s4d$\,$^3\!D_1$ &  3655  &  2.318(5)  &    -2       &  -1     &   -3      \\
$5s4d$\,$^3\!D_2$ &  3714  &  4.013(9)  &    -6       &   1     &   -5     \\
$5s6s$\,$^3\!S_1$ &  14534 &  3.435(14) &   -27       & -13     &   -40     \\
$5s5d$\,$^3\!D_1$ &  20503 &  2.005(20) &   -38(1)    & -19     &  -57    \\
$5s5d$\,$^3\!D_2$ &  20518 &  3.671(36) &   -128(3)   &  13     &  -115     \\
$5p^2$\,$^3\!P_0$ &  20689 &  2.658(27) &   -72(1)    &  72(1)  &    0     \\
$5p^2$\,$^3\!P_1$ &  20896 &  2.363(24) &   -63(1)    & -32(1)  &  -95       \\
$5p^2$\,$^3\!P_2$ &  21170 &  2.867(29) &   -108(2)   & 11      &  -97    \\
$5p^2$\,$^1\!D_2$ &  22457 &  0.230(11) &   -3        &  0      &   -3        \\
$5p^2$\,$^1\!S_0$ &  22656 &  0.301(17) &   -8(1)     &  8(1)   &   0      \\
$5s7s$\,$^3\!S_1$ &  22920 &  0.922(28) &   -842(51)  & -421(25)&   -1263      \\
$5s6d$\,$^3\!D_1$ &  25181 &  1.025(9) &     12       &    6    &   18      \\
$5s6d$\,$^3\!D_2$ &  25186 &  1.730(12) &    34       &   -3    &   31      \\
Other            &        &            &    66(7)     &   -1    &  65      \\
Core+vc          &        &            &     6        &         &  6        \\
Total            &        &            &   -1178(51)  & -380(25)& -1558(57)
   \end{tabular}
\end{ruledtabular}
\end{table}
This calculation is illustrated in Table ~\ref{tab2t} where we list contributions to the $5s5p\,^3\!P_1$ polarizability in atomic units at the 435.827~nm $^1\!S_0 - ^3\!P_1$ magic wavelength.
The scalar polarizability has a small contribution from polarizability of the ionic Sr$^{2+}$ core and a ``vc'' term that
compensates for a Pauli principle violating core-valence excitation from the core to the valence shells. It is calculated using the random phase approximation (RPA) \cite{Sr2013} and listed in row labelled ``Core+vc''. Since the core has zero angular momentum, there is no core contribution to the tensor polarizability. 
Transition energies $\Delta E$ in cm$^{-1}$ \cite{NIST} and the recommended values of the electric-dipole matrix elements $D$ from the compilation of ~\cite{Kestler2022} or the CI+all-order calculation are listed in atomic units. 

The breakdown of the contributions shows that the $^3\!P_1$ polarizability is strongly dominated by the contributions from the $5s5p \,^3\!P_1 - 5s7s\, ^3\!S_1$ transition.
This was not the case in ~\cite{Kestler2022} where the contributions of the transitions to  $5p^2\, ^3\!P_J$ states strongly dominated.
\begin{figure}[t!]
    \centering
    \includegraphics[width=\columnwidth]{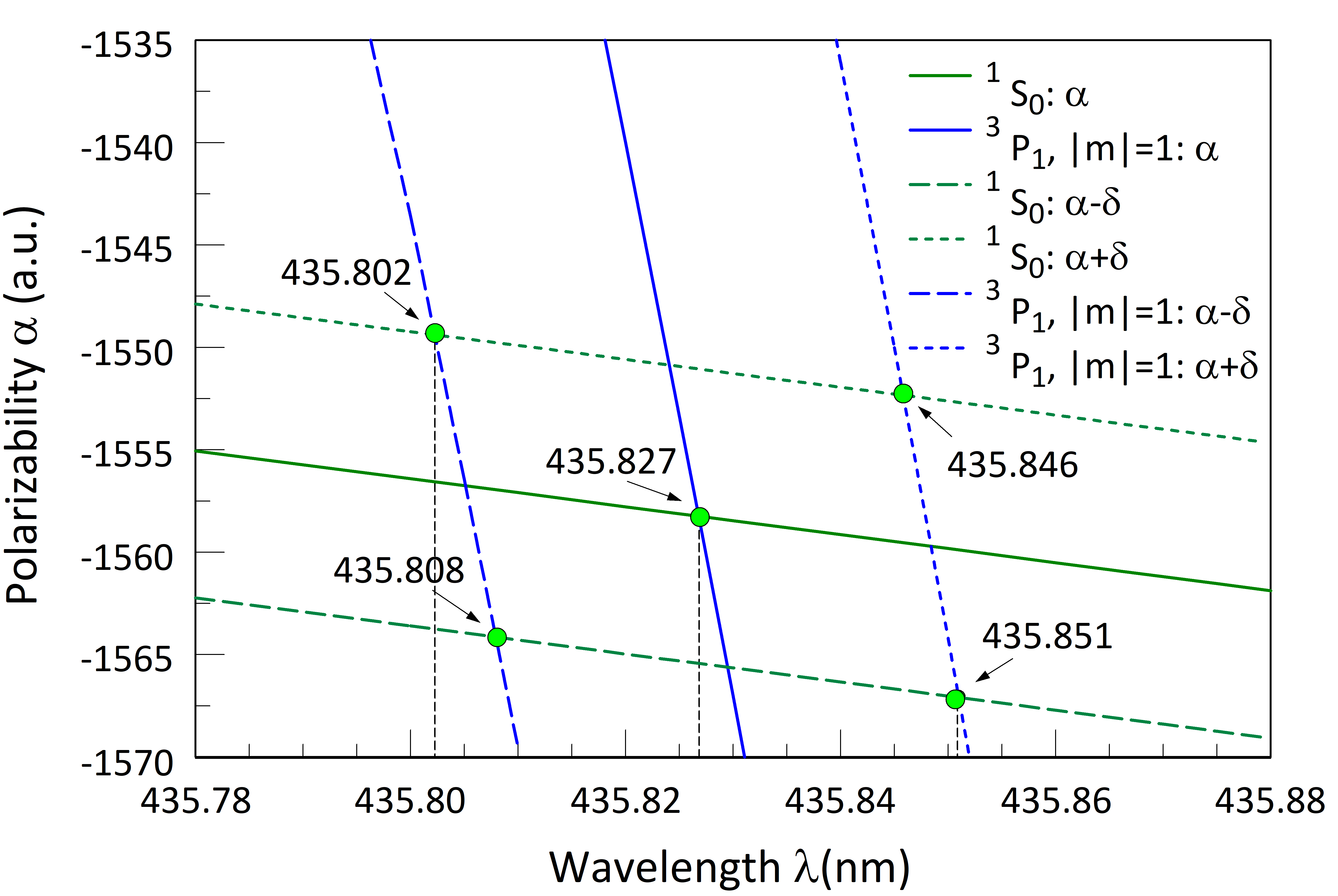}
    \caption{Calculated polarizabilities for the $5s^{2}\,^1\!S_{0}$ and $5s5p\,^3\!P_{1,|m|=1}$ levels and determination of the magic wavelength.}
    \label{fig:pol_calc}
\end{figure}

To evaluate the uncertainties of the theory matrix elements, we carry out another calculation using a method that combines CI and second-order MBPT (many-body perturbation theory). Since the treatment of the two valence electrons is essentially complete in the CI method, the main source of the uncertainty is the treatment of the core and core-valence correlations included via the effective Hamiltonian. 
The CI+MBPT method does not include higher-order corrections to the effective Hamiltonian, and the difference of the matrix elements calculated with CI+ All-order and MBPT techniques gives the size of the dominant higher-order corrections. We take the full size of the higher-order correction as an estimate of the uncertainty. This method has been tested on the computation of matrix elements that are experimentally known. The uncertainty of polarizability comes from the uncertainties in the matrix elements and the small ``Other'' remainder.  At the magic wavelength, the entire uncertainty comes from the uncertainty in the $5s5p \,^3\!P_1 - 5s7s\, ^3\!S_1$ matrix elements as illustrated in Table~\ref{tab2t}.

The determination of the magic wavelength and its uncertainty is illustrated in Fig.~\ref{fig:pol_calc},
where we plot the calculated polarizabilities for the $5s^{2}\,^1\!S_{0}$ and $5s5p\,^3\!P_{1}$, $|m|=1$ states. Magic wavelength is  obtained as the crossing points. We plot $\alpha+\delta \alpha$ and $\alpha-\delta \alpha$, where $\delta \alpha $ is the uncertainty on the polarizability.  The uncertainty in the value of the magic wavelength is determined from the crossings of these additional curves. 

Finally, regarding the calculation of the experimentally observed shift of the $^{1}\!S_{0}-{}^{1}\!P_{1}$ transition, it is instructive to first note that there are several major classes of methods used to determine atomic wave functions and energy levels that can be adapted to generate polarizabilities. Besides the Configuration Interaction (CI), there are Density Functional Theory (DFT), Correlated Basis Function methods, Coupled Cluster (CC) methods, Many-Body Perturbation Theory (MBPT), and many of their different types and implementations. Each of these methods has its own features, scope and shortages. The CI+all-orders method (a combination of CI and linearized CC methods) is one of the most powerful and accurate methods as it includes  higher-order corrections to a much larger extent than other approaches. To the best of our knowledge, the dynamic polarizability in the considered part of the wavelengths for the ${}^{1}\!P_{1}$ state of Sr I has not been previously calculated by any other methods. The problem mentioned in the manuscript is a numerical issue that appears to originate from the use of B-spline basis sets built in large in the configurational interaction calculations, not present for smaller cavity sizes. A large cavity is required to correctly reproduce properties of higher lying states, such as 5s11d  that give a large contribution  to ${}^{1}\!P_{1}$ polarizability at 435.8 nm.  We have done extensive numerical tests that led us to believe solving this problem likely requires introducing different basis sets of modifying the grid that is beyond the scope of the present work.

\subsection{\label{app:blue_magic_exp}Experimental Shift}

The magic wavelength measurements and theoretical calculations are strongly dependent on proper field polarization settings. Given the low visibilities reported in Appendix~\ref{app:pol_calibration}, we propagate these errors as a systematic uncertainty to our trapping field. Unable to reach purely horizontal (vertical) polarization defines a possible range of magic wavelengths given the theoretical polarizabilities.

The evanescent trapping fields can be written as a combination of the right hand circular ${\bf{E}}^{+}$ and and left hand circular ${\bf{E}}^{-}$ fields as
$${\bf{E}}(\phi, \theta) = \cos(\theta/2){\bf{E}}^{+} + e^{-i\phi}\sin(\theta/2){\bf{E}}^{-},$$
where $\phi$ and $\theta$ are defined on the Poincar\'e sphere to be the azimuthal and elevation angles. Given the polarization uncertainty measurements described above, the circle around purely horizontal (vertical) polarization is parameterized by a single angle $\theta_{\rm{err}}$, thus $\phi = \phi(\theta_{\rm{err}})$.

\begin{table}[t!]
\caption{\label{tab:supp_polarizabilities} Theoretical atomic polarizabilities of the ${}^{1}S_{0}$ and ${}^{3}P_{1,|m|=1}$ states in a.u. for the two trapping wavelengths. The singlet ${}^{1}S_{0}$ only has a scalar contribution $\alpha_{s}$, while the triplet state contributes scalar, vector, and tensor polarizabilities ($\alpha_{s}, \alpha_{v}, \alpha_{t}$), which are used to determine the total energy shift caused by the evanescent fields.}
{\renewcommand{\arraystretch}{1.2}
\begin{tabular}{p{2.0cm}>{\centering}p{0.0cm}>{\centering}p{1.15cm}>{\centering}p{0.5cm}>{\centering}p{1.5cm}>{\centering}p{1.5cm}>{\centering\arraybackslash}p{1.15cm}}
&&\multicolumn{1}{c}{${}^{1}S_{0}$}&&\multicolumn{3}{c}{${}^{3}P_{1,|m|=1}$}\\\cline{2-3}\cline{5-7}Wavelength&&$\alpha_{s}$&&$\alpha_{s}$&$\alpha_{v}$&$\alpha_{t}$\\
\hline
435.97 && $-1570.14$ && $-1555.57$ & $1925.03$ & $-567.66$ \\
473.251 && $3651.76$ && $3444.04$ & $9051.92$ & $-442.87$ \\
\hline\hline
\end{tabular}}
\end{table}

\begin{figure}[t!]
    \includegraphics{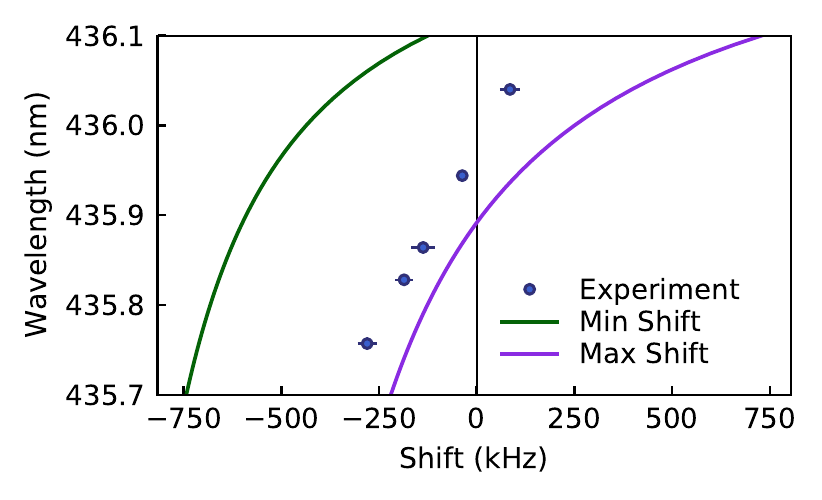}
    \caption{Total energy shift of the ${}^{1}S_{0} - {}^{3}P_{1}$ transition in the lattice trap. For different configurations around the polarization uncertainty region [red circle on Fig.~\ref{fig:pol_setup}(c)] for each trapping beam, the minimum (maximum) energy shift at various wavelengths is illustrated with the green (purple) line.} 
    \label{fig:magic_wavelength_uncertainties}
\end{figure}

Given the theoretical atomic polarizabilities in Table~\ref{tab:supp_polarizabilities} and the evanescent field, we calculate the total frequency shift of the ${}^{1}S_{0}-{}^{3}P_{1}$ transition as
\begin{align*}
    h\Delta f = -\Delta\alpha_{s}|{\bf{E}}(\theta)|^{2} &+ \alpha_{v}(i{\bf{E}}^{\dag}(\theta) \times {\bf{E}}(\theta)) \nonumber\\
    &- \alpha_{t}\frac{\bigg(3|{\bf{E}}(\theta)_{y}|^{2} - |{\bf{E}}(\theta)|^{2}\bigg)}{2},
\end{align*}
where ${\bf{E}}(\theta)_{y}$ is the field component along the quantization axis $\hat{y}$ and $\Delta\alpha_{s}$ is the difference of the ${}^{1}S_{0}$ and ${}^{3}P_{1}$ scalar polarizabilities $\alpha_{s}$ in Table~\ref{tab:supp_polarizabilities}. Maximizing (minimizing) the energy shift with respect to the polarization angle $\theta$ around the circle on the Poincar\'e sphere, yields a range of possible zero crossings of the frequency shift (see Fig.~\ref{fig:magic_wavelength_uncertainties}). The minimum energy shift occurs when the blue-detuned field is completely linearly polarized, but diagonal, while one red-detuned beam is completely linearly polarized and the other is completely circularly polarized. The maximum energy shift occurs with oppositely polarized beams so that the blue-detuned field is completely circularly polarized and the two red-detuned beams also switch polarization.

\section{\label{app:shelving_calibration}Shelving Calibration}

The shelving beam was calibrated by free-space absorption imaging on the strong ${}^{1}S_{0}-{}^{1}P_{1}$, 461-nm transition. The external shelving beam is aligned in the downwards direction while the magnetic field sets the quantization axis to be in the upwards direction. The polarization of the shelving beam is set to $\sigma^{-}$ to maximize transfer of the ${}^{3}P_{1, m=-1}$ state on resonance. The shelving beam, measured at 6~cm from the atoms, has a power of $\approx$200~$\mu$W and a beam waist of $w_{0}\approx1.5$~mm.

\begin{figure}[!t]
    \includegraphics{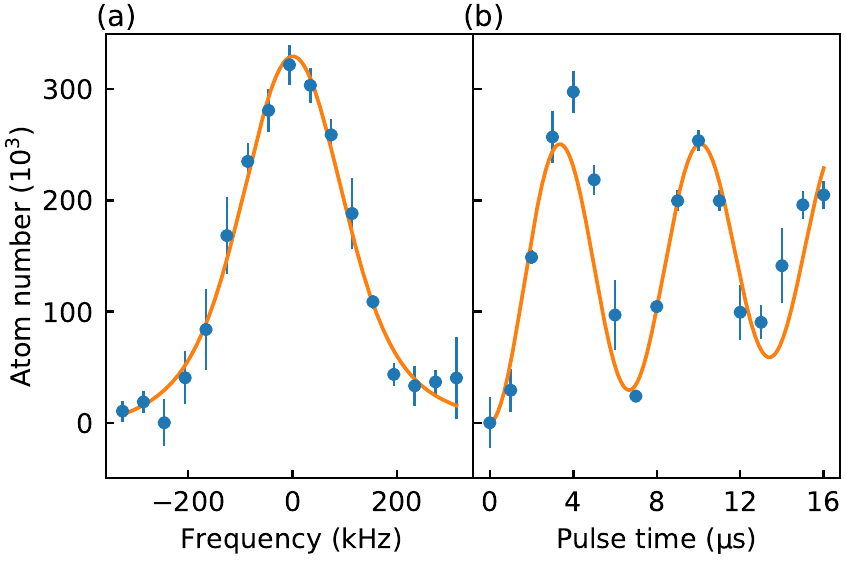}
    \caption{Calibration of the shelving beam for the narrow-line intercombination transition ${}^{1}S_{0}-{}^{3}P_{1, m=-1}$ with the MOT cloud at $3$~ms time-of-flight. (a) Resonance of the transition by shelving spectroscopy. We measure the number of atoms in the excited state $^{3}P_{1,m=-1}$ transferred by the shelving beam. (b) Rabi frequency calibrated by varying the shelving beam pulse time.}
    \label{fig:shelving_calibration}
\end{figure}

We first perform shelving spectroscopy to find the ${}^{1}S_{0}-{}^{3}P_{1, m=-1}$ resonance before varying the shelving pulse time to measure the Rabi frequency $\Omega_{R}=2\pi\times148(3)$~kHz consistent with a beam intensity of 2.3~mW/cm$^{2}$, as shown in Fig.~\ref{fig:shelving_calibration}. The spectral linewidth is broadened from the natural $\Gamma=2\pi\times7.6$~kHz to $\Delta\omega_{\rm{FWHM}} = 2\pi\times240$~kHz by power broadening ($\approx$200~kHz) and thermal Doppler broadening ($\approx$40~kHz)

With a shelving pulse time of $4~\mu$s, we transfer $\approx$70\% of the cold atom population to the excited ${}^{3}P_{1,m=-1}$ state, consistent with the on-resonance absorption measurements during the shelving spectroscopy through the nanofiber shown in Fig. ~\ref{fig:shelving_spectroscopy}(a). With the low probing power of 50(3)~pW, shorter pulse lengths, $\ll21$~$\mu$s, limit the signal recorded by the SPCM while longer pulse lengths, $\approx$21~$\mu$s, allow atoms to decay back to the ground state and can contribute to the absorption signal. We empirically determined the pulse length of 15~$\mu$s to be optimal.

\end{appendix}

\bibliography{references}

\end{document}